\documentclass[10pt]{article}
\usepackage{graphicx}	

\usepackage{scicite}

\usepackage{times}

\newcommand\arcsec{\mbox{$^{\prime\prime}$}}%
\newcommand{\beginsupplement}{%
        \setcounter{table}{0}
        \renewcommand{\thetable}{S\arabic{table}}%
        \setcounter{figure}{0}
        \renewcommand{\thefigure}{S\arabic{figure}}%
     }
\usepackage{float}
\floatstyle{plaintop}
\restylefloat{table}

\newenvironment{sciabstract}{%
\begin{quote} \bf}
{\end{quote}}

\title{{\it Swift} and {\it NuSTAR} observations of GW170817: detection of a blue kilonova}

\date{}

\begin{document} 


\baselineskip14pt


\maketitle

{
\Large
\noindent
P.A. Evans$^{1\ast}$, 
S.B. Cenko$^{2,3}$, 
J.A. Kennea$^{4}$, 
S.W.K. Emery$^{5}$, \and
N.P.M. Kuin$^{5}$, 
O. Korobkin$^{6}$, 
R.T. Wollaeger$^{6}$, 
C.L. Fryer$^{6}$, 
K.K. Madsen$^{7}$, \and
F.A. Harrison$^{7}$, 
Y. Xu$^{7}$, 
E. Nakar$^{8}$, 
K. Hotokezaka$^{9}$, 
A. Lien$^{10,11}$, 
S. Campana$^{12}$, \and
S.R. Oates$^{13}$, 
E. Troja$^{2,14}$, 
A.A. Breeveld$^{5}$, 
F. E. Marshall$^{2}$, 
S.D. Barthelmy$^{2}$, \and
A. P. Beardmore$^{1}$, 
D.N. Burrows$^{4}$, 
G. Cusumano$^{15}$, 
A. D'A\`\i$^{15}$, 
P. D'Avanzo$^{12}$, \and
V. D'Elia$^{16,17}$, 
M. de~Pasquale$^{18}$, 
W.P. Even$^{6,19}$, 
C.J. Fontes$^{6}$, 
K. Forster$^{7}$, \and
J. Garcia$^{7}$, 
P. Giommi$^{17}$, 
B. Grefenstette$^{7}$, 
C. Gronwall$^{4,20}$, 
D.H. Hartmann$^{21}$, \and
M. Heida$^{7}$, 
A.L. Hungerford$^{6}$, 
M.M. Kasliwal$^{22}$, 
H.A. Krimm$^{23,24}$, 
A.J. Levan$^{13}$, \and
D. Malesani$^{25}$, 
A. Melandri$^{12}$, 
H. Miyasaka$^{7}$, 
J.A. Nousek$^{4}$, 
P.T. O'Brien$^{1}$, \and
J.P. Osborne$^{1}$, 
C. Pagani$^{1}$, 
K.L. Page$^{1}$, 
D.M. Palmer$^{26}$, 
M. Perri$^{16,17}$, 
S. Pike$^{7}$, \and
J.L. Racusin$^{2}$, 
S. Rosswog$^{27}$, 
M.H. Siegel$^{4}$, 
T. Sakamoto$^{28}$, 
B. Sbarufatti$^{4}$, \and
G. Tagliaferri$^{12}$, 
N.R. Tanvir$^{1}$, 
A. Tohuvavohu$^{4}$ \and

\vspace{1cm}
\noindent\normalsize{$^{1}$ University of Leicester, X-ray and Observational Astronomy Research Group, Leicester Institute for Space and Earth Observation, Department of Physics \& Astronomy, University Road, Leicester, LE1 7RH, UK} \and

\noindent\normalsize{$^{2}$ Astrophysics Science Division, NASA Goddard Space Flight Center, Greenbelt MD, 20771 USA} \and

\noindent\normalsize{$^{3}$ Joint Space-Science Institute, University of Maryland, College Park, MD 20742, USA} \and

\noindent\normalsize{$^{4}$ Department of Astronomy and Astrophysics, The Pennsylvania State University, University Park, PA 16802, USA} \and

\noindent\normalsize{$^{5}$ University College London, Mullard Space Science Laboratory, Holmbury St. Mary, Dorking, RH5 6NT, U.K.} \and

\noindent\normalsize{$^{6}$ Center for Theoretical Astrophysics, Los Alamos National Laboratory, Los Alamos, NM 87545 USA} \and

\noindent\normalsize{$^{7}$ Cahill Center for Astronomy and Astrophysics, California Institute of Technology, 1200 East California Boulevard, Pasadena, CA 91125, U} \and

\noindent\normalsize{$^{8}$ The Raymond and Beverly Sackler School of Physics and Astronomy, Tel Aviv University, Tel Aviv 69978, Israel} \and

\noindent\normalsize{$^{9}$ Center for Computational Astrophysics, Simons Foundation, 162 5th Ave, New York, 10010, NY, USA} \and

\noindent\normalsize{$^{10}$ Center for Research and Exploration in Space Science and Technology (CRESST) and NASA Goddard Space Flight Center, Greenbelt MD, 20771 USA} \and

\noindent\normalsize{$^{11}$ Department of Physics, University of Maryland, Baltimore County, 1000 Hilltop Circle, Baltimore, MD 21250, USA} \and

\noindent\normalsize{$^{12}$ INAF -- Osservatorio Astronomico di Brera, Via Bianchi 46, I-23807 Merate, Italy} \and

\noindent\normalsize{$^{13}$ Department of Physics, University of Warwick, Coventry, CV4 7AL, UK} \and

\noindent\normalsize{$^{14}$ Department of Physics and Astronomy, University of Maryland, College Park, MD 20742-4111, USA} \and

\noindent\normalsize{$^{15}$ INAF -- IASF Palermo, via Ugo La Malfa 153, I-90146, Palermo, Italy} \and

\noindent\normalsize{$^{16}$ INAF-Osservatorio Astronomico di Roma, via Frascati 33, I-00040 Monteporzio Catone, Italy} \and

\noindent\normalsize{$^{17}$ Space Science Data Center (SSDC) - Agenzia Spaziale Italiana (ASI), I-00133 Roma, Italy } \and

\noindent\normalsize{$^{18}$ Department of Astronomy and Space Sciences, University of Istanbul, Beyz{\i}t 34119, Istanbul, Turkey} \and

\noindent\normalsize{$^{19}$ Southern Utah University, Cedar City, UT 84720, USA} \and

\noindent\normalsize{$^{20}$ Institute for Gravitation and the Cosmos, The Pennsylvania State University, University Park, PA 16802} \and

\noindent\normalsize{$^{21}$ Department of Physics and Astronomy, Clemson University, Kinard Lab of Physics, USA} \and

\noindent\normalsize{$^{22}$ Division of Physics, Mathematics and Astronomy, California Institute of Technology, Pasadena, CA 91125, USA} \and

\noindent\normalsize{$^{23}$ Universities Space Research Association, 7178 Columbia Gateway Dr, Columbia, MD 21046, USA} \and

\noindent\normalsize{$^{24}$ National Science Foundation, 2415 Eisenhower Avenue, Alexandria, VA 22314, USA} \and

\noindent\normalsize{$^{25}$ Dark Cosmology Centre, Niels Bohr Institute, University of Copenhagen, Juliane Maries Vej 30, DK-2100 Copenhagen \O, Denmark} \and

\noindent\normalsize{$^{26}$ Los Alamos National Laboratory, B244, Los Alamos, NM, 87545, USA } \and

\noindent\normalsize{$^{27}$ The Oskar Klein Centre, Department of Astronomy, AlbaNova, Stockholm University, SE-106 91 Stockholm, Sweden} \and

\noindent\normalsize{$^{28}$ Department of Physics and Mathematics, Aoyama Gakuin University, Sagamihara, Kanagawa, 252-5258, Japan} \and
\\
\noindent\normalsize{$^\ast$To whom correspondence should be addressed; E-mail:  pae9@leicester.ac.uk}
}


\begin{sciabstract}
With the first direct detection of merging black holes in 2015, the 
era of gravitational wave (GW) astrophysics began.  A complete
picture of compact object mergers, however, requires the detection
of an electromagnetic (EM) counterpart. We report ultraviolet (UV) 
and X-ray observations by {\it Swift} and the {\it Nuclear Spectroscopic 
Telescope ARray} ({\it NuSTAR}) of the EM counterpart of the binary 
neutron star merger GW\,170817. The bright, rapidly fading ultraviolet 
emission indicates a high mass ($\approx0.03$ solar masses) wind-driven 
outflow with moderate electron fraction ($Y_{e}\approx0.27$).  
Combined with the X-ray limits, we favor an observer viewing angle
of $\approx 30^{\circ}$ away from the orbital rotation axis, which 
avoids both obscuration from the heaviest elements in the orbital 
plane and a direct view of any ultra-relativistic, highly collimated
ejecta (a gamma-ray burst afterglow).

 \end{sciabstract}

\section*{One-sentence summary}
We report X-ray and UV observations of the first binary neutron star merger 
detected via gravitational waves.


\section*{Main Text}

At 12:41:04.45 on 2017 August 17 (UT times are used throughout this work), the
Laser Interferometric Gravitational-Wave Observatory (LIGO) and Virgo 
Consortium (LVC) registered a strong gravitational wave (GW) 
signal (LVC trigger G298048; \cite{LVCC21505}), later named 
GW\,170817 \cite{LVCDiscovery}.  Unlike previous GW sources reported by 
LIGO, which involved only black holes \cite{Abbott16X}, the gravitational 
strain waveforms indicated a merger of two neutron stars.  Binary neutron star 
mergers have long been considered a promising candidate for the 
detection of an electromagnetic counterpart associated with a gravitational 
wave source.

Two seconds later, the Gamma-Ray Burst Monitor (GBM) on the \emph{Fermi} spacecraft triggered on a short (duration
$\approx 2$\,s) gamma-ray signal consistent with the GW localization, GRB\,170817A \cite{LVCC21506,Goldstein17}. The location of the
{\it Swift} satellite \cite{GehrelsSwift} in its low-Earth orbit meant that the GW and gamma-ray burst (GRB)
localizations were occulted by the Earth \cite{E17SM} and so not visible to its Burst Alert Telescope (BAT). These
discoveries triggered a world-wide effort to find, localize and characterize the EM counterpart \cite{Capstone}. We
present UV and X-ray observations conducted as part of this campaign;  companion papers describe synergistic
efforts at radio \cite{Hallinan17} and optical/near-infrared \cite{Kasliwal17} wavelengths.

\section*{Search for a UV and X-ray Counterpart}

\begin{figure}
\begin{center}
\includegraphics[width=12cm]{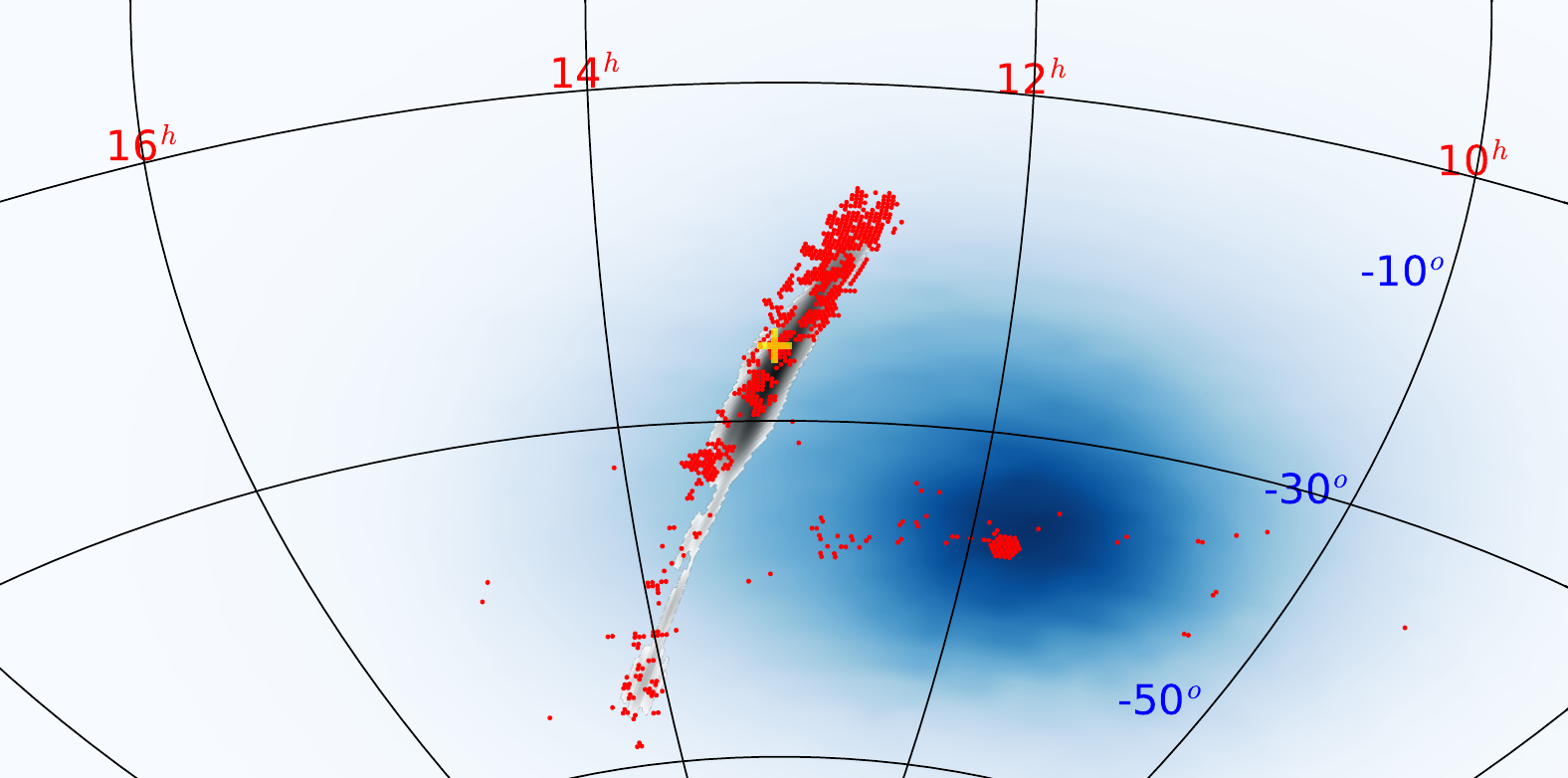}
\end{center}

\caption{{\bf Skymap of \emph{Swift} XRT observations, in equatorial (J2000) coordinates}. The grey
probability area is the GW localization \cite{LVCC21527},
the blue region shows the \emph{Fermi}-GBM localization,  
and the red circles are {\it Swift}-XRT fields of view.
UVOT fields are colocated with a field of view 60\%\ of the XRT.
The location of the counterpart, EM\,170817, is marked with a large yellow cross.
The early 37-point mosaic can be seen, centred on the GBM probability. 
The widely scattered points are from the first uploaded observing plan, 
which was based on the single-detector GW skymap.
The final observed plan was based on the first 3-detector map \cite{LVCC21513}, however
we show here the higher-quality map \cite{LVCC21527} so that our coverage can 
be compared to the final probability map (which was not available at 
the time of our planning; \cite{E17SM}).
}
\label{fig:skymap}
\end{figure}

\emph{Swift} began searching for a counterpart to GW\,170817 with its X-ray Telescope (XRT) and UV/Optical Telescope
(UVOT) at 13:37 (time since the GW and GRB triggers, $\Delta t = 0.039$\,d). At the time, the most precise
localization was from the \emph{Fermi}-GBM (90\% containment area of 1626\,deg$^{2}$), so we imaged a mosaic with
radius $\sim1.1^{\circ}$ centered on the most probable GBM position. Subsequently at 17:54 ($\Delta t = 0.2$\,d) a more
precise localization became available from the LIGO and Virgo GW detectors, with a 90\% containment area of only
33.6\,deg$^{2}$ \cite{LVCC21513}. Following the strategy outlined in \cite{Evans16c}, \textit{Swift} began a series of
short (120\,s) exposures centered on known galaxies in the GW localization (Figure~\ref{fig:skymap}; \cite{E17SM}).

No new, bright (X-ray flux, $f_{X} \geq 10^{-12}$\,erg\,cm$^{-2}$\,s$^{-1}$) X-ray sources
were detected in the wide-area search (XRT imaged 92\% of the distance-weighted
GW localization \cite{E17SM}). 
In order to quantify the likelihood of recovering any rapidly
fading X-ray emission, we simulated 10,000 short GRB afterglows based on a
flux-limited sample of short GRBs \cite{Davanzo14}, and randomly placed them in the 3D
(distance plus sky position) GW localization, weighted by the GW probability. We
find that in 65\% of these simulations we could recover an X-ray afterglow with
our wide-area tiling observations \cite{E17SM}.

At 01:05 on 2017 August 18 ($\Delta t=0.5$\,d), 
a candidate optical counterpart, Swope Supernova Survey 17a (SSS17a) 
\cite{LVCC21529,Coulter17}, was reported in the galaxy NGC~4993 (distance $d \approx
40$\,Mpc).  Ultimately this source, which we refer to
as EM\,170817, was confirmed as the electromagnetic
counterpart to the GW detection and the \emph{Fermi}
GRB \cite{Capstone}, making it the closest known short GRB to Earth.  Follow-up
observations of EM\,170817 \cite{E17SM} with \emph{Swift} began at 
03:34 ($\Delta t = 0.6$\,d) and with the \emph{Nuclear Spectroscopic 
Telescope ARray (NuSTAR)} \cite{Harrison13} at 05:25 ($\Delta t = 0.7$\,d).

\begin{figure}
\begin{center}
\includegraphics[width=18cm]{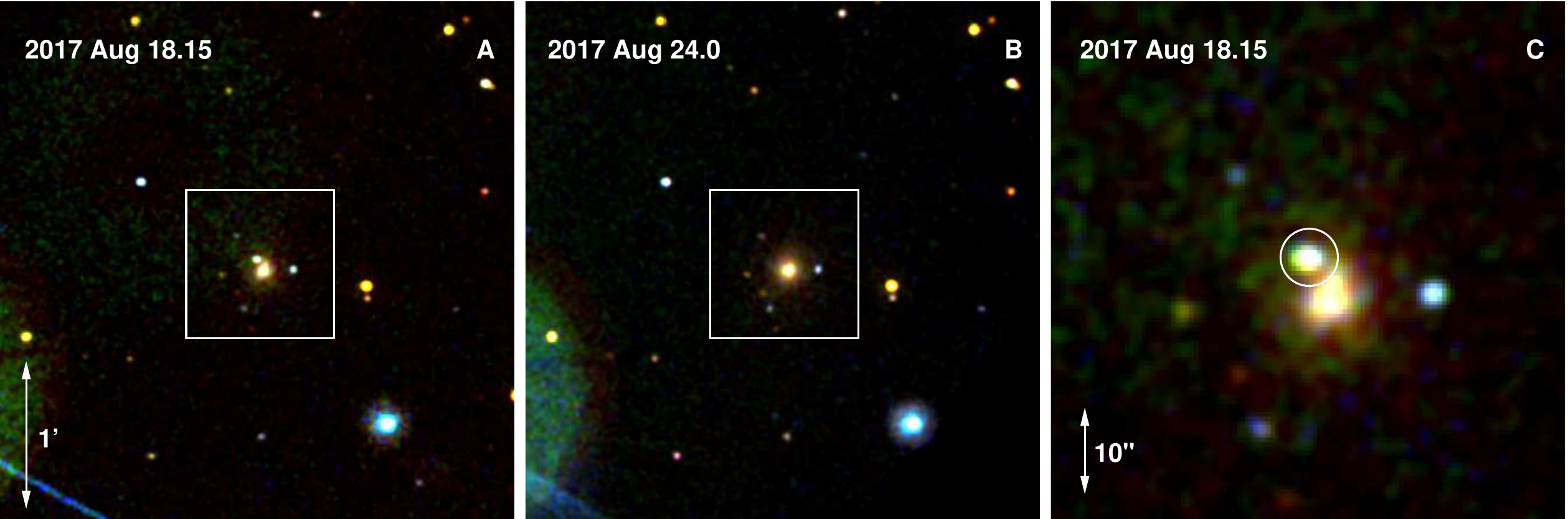}
\end{center}
\caption{{\bf False-color UV image of the field of EM\,170817}.
The $u$, $uvw1$ and $uvm2$ filters have been assigned to the red, green and blue channels
respectively.
Bright UV emission is clearly detected in our first epoch (panel A),
which rapidly fades at blue wavelengths (panel B).  Panel C
shows a zoom in of the first epoch with the transient circled.  All images are oriented with
North up and East to the left.}
\label{fig:uvotfinder}
\end{figure}

\begin{figure}
\begin{center}
\includegraphics[width=10cm]{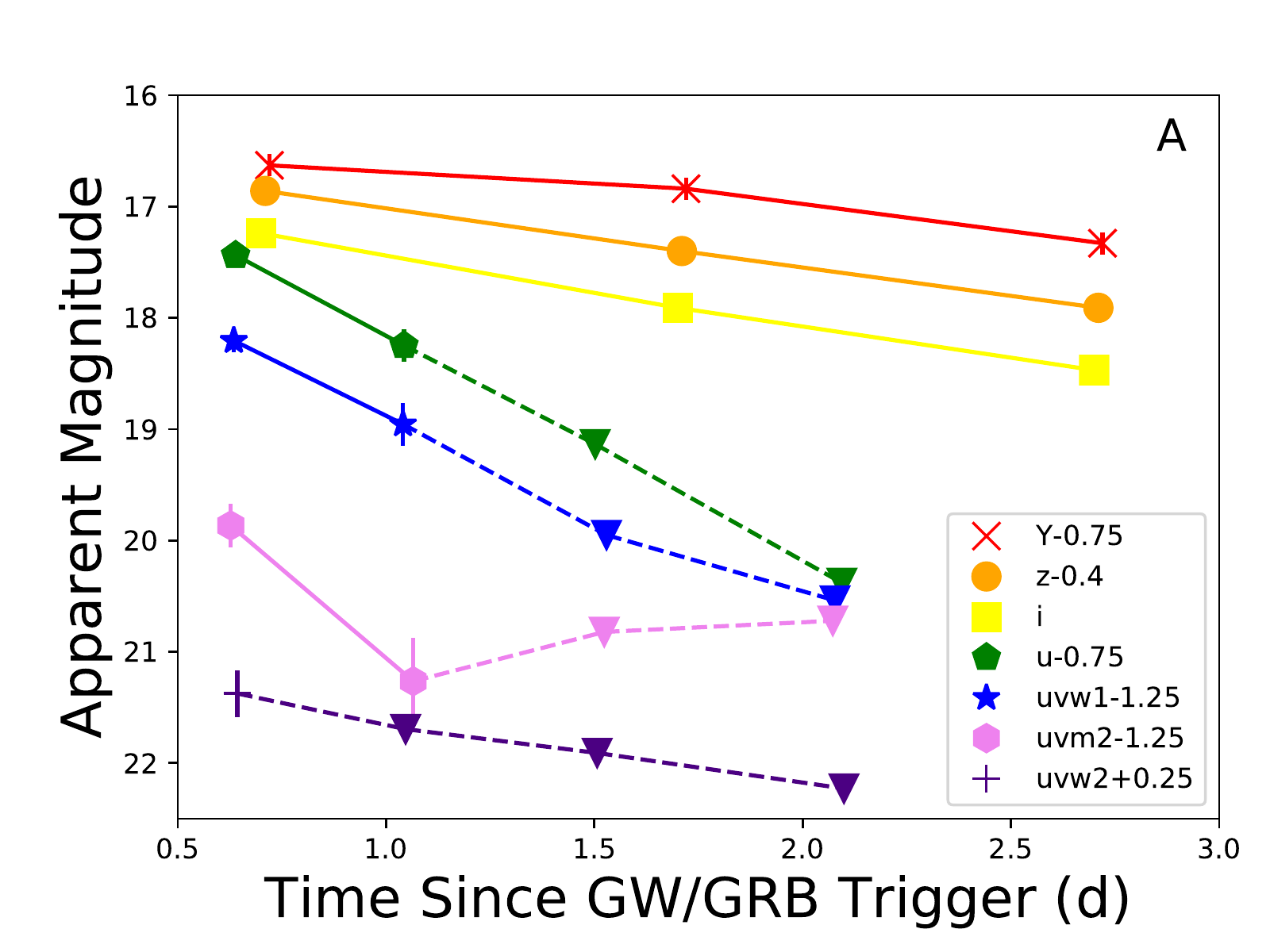}
\includegraphics[width=15cm]{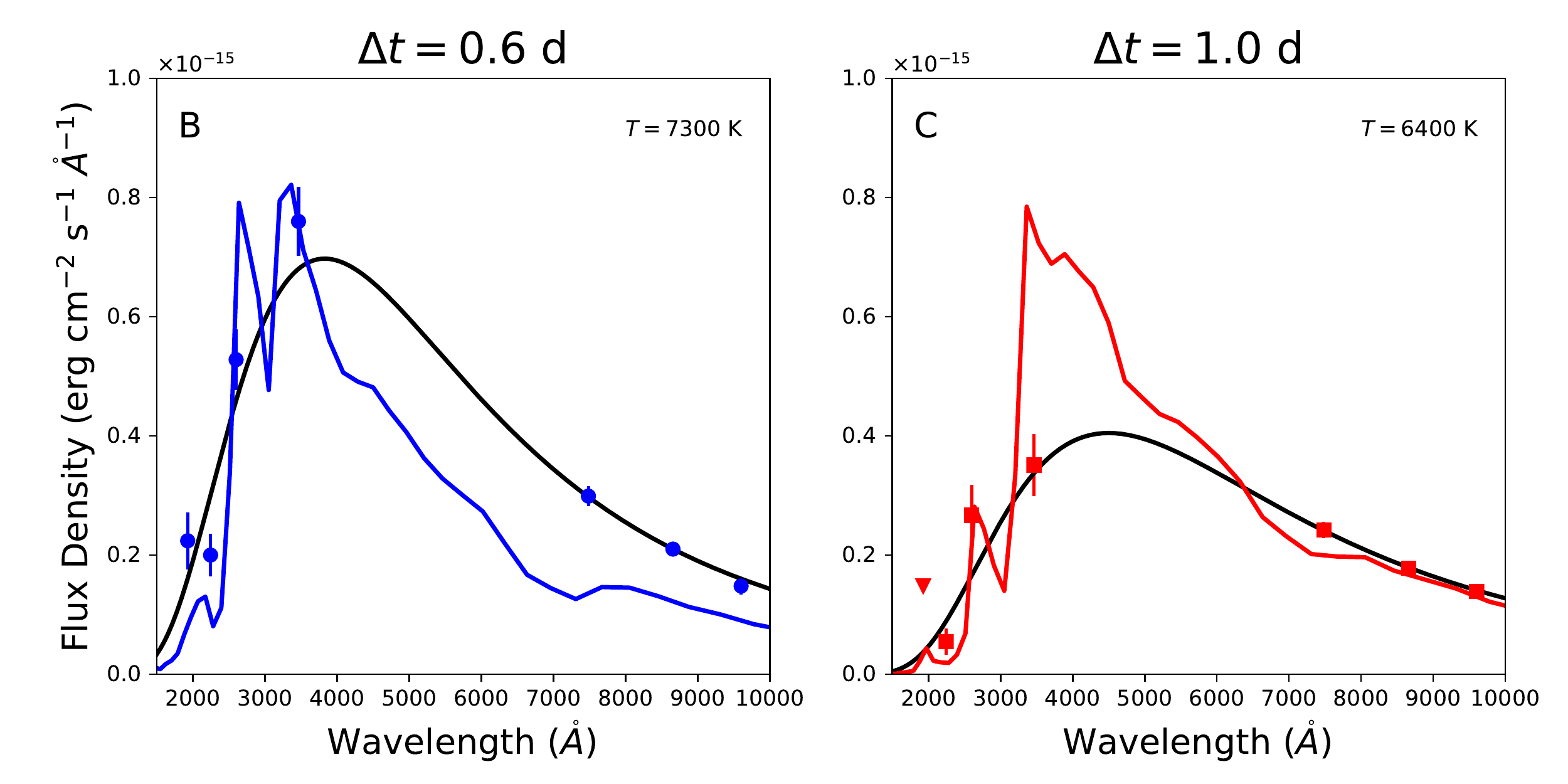}
\end{center}
\caption{{\bf UV and optical light curves and SEDs.} \emph{A:} \emph{Swift}-UVOT light curve of the optical counterpart
EM\,170817 of GW\,170817.  The data are corrected for host 
galaxy contamination.  Upper limits are plotted as inverted triangles.
Also shown are host-subtracted optical and near-infrared
photometry from Pan-STARRS \cite{Smartt17}.  
\emph{B-C} The spectral energy distribution of EM\,170817, with blackbody models (black curves) demonstrating the rapid cooling of the ejecta.
Overplotted are the best fitting kilonova models (colored lines), where the wind ejecta have mass
$0.03$\,$M_{\odot}$, and velocity $0.08c$, while the dynamical ejecta have mass $0.013$\,$M_{\odot}$ and
velocity  $0.3c$ \cite{E17SM}. The red triangle in the right hand figure is a 3-$\sigma$ upper limit.}
\label{fig:uv}
\end{figure}

In the first exposures ($\Delta t = 0.6$\,d), the UVOT detected a bright 
fading UV source at the location of EM\,170817 (Figure~\ref{fig:uvotfinder}).
The initial magnitude was $u = 18.19_{-0.08}^{+0.09}$\, mag (AB), but 
subsequent exposures revealed rapid fading at UV wavelengths.
The rapid decline in the UV is in  contrast to the optical and near-infrared 
emission, which remained flat for a much longer period of time 
(Figure~\ref{fig:uv}; \cite{Kasliwal17}).  

Neither the \emph{Swift}-XRT nor \emph{NuSTAR} instruments detected
X-ray emission at the location of EM\,170817.  A full listing of the
\emph{Swift}-XRT and {\it NuSTAR} upper limits at this location is provided 
in Table S2.

\section*{The UV Counterpart Rules Out an On-Axis Afterglow}

\begin{figure}
\begin{center}
\includegraphics[width=10cm]{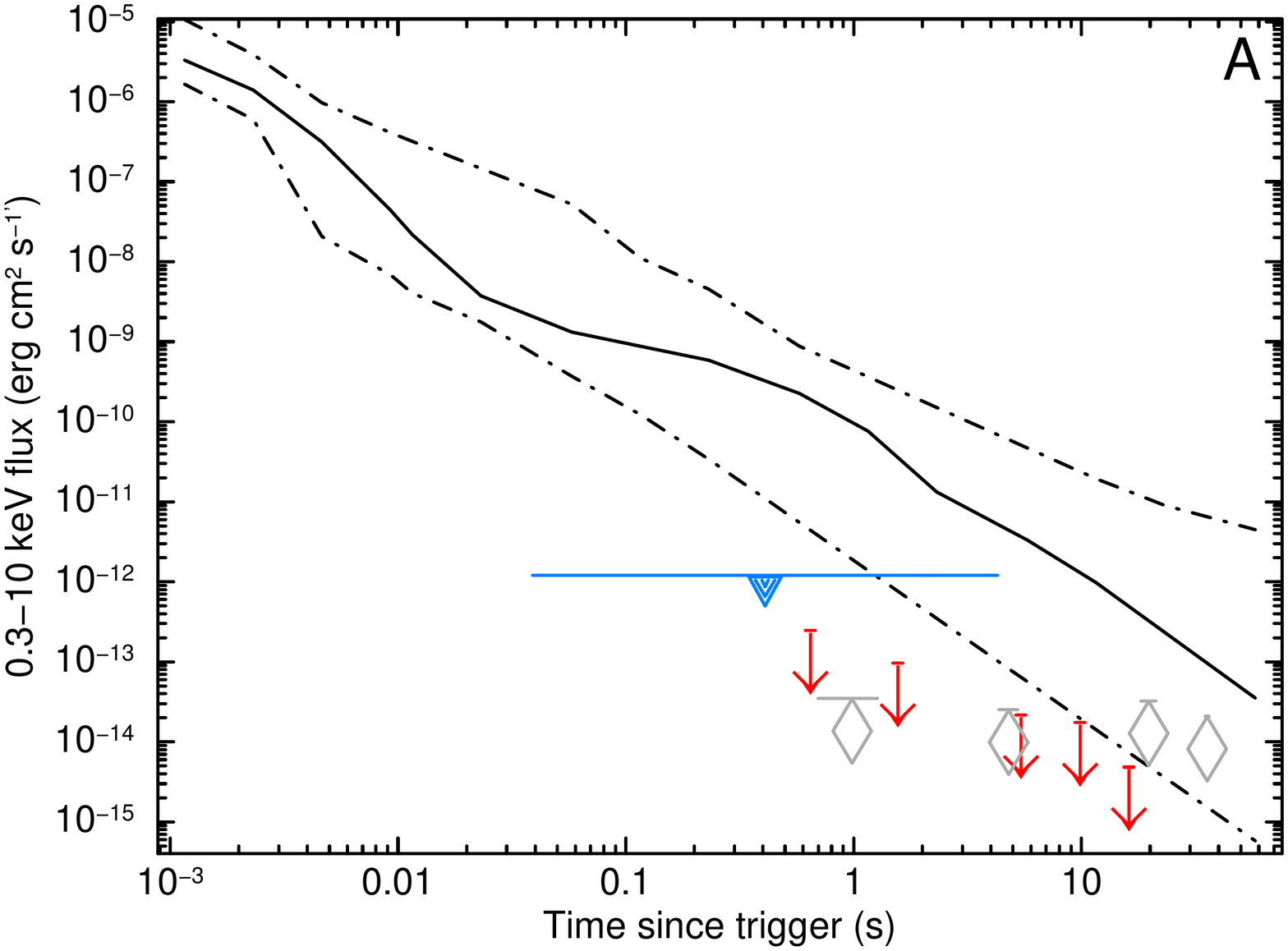}
\includegraphics[width=10cm,]{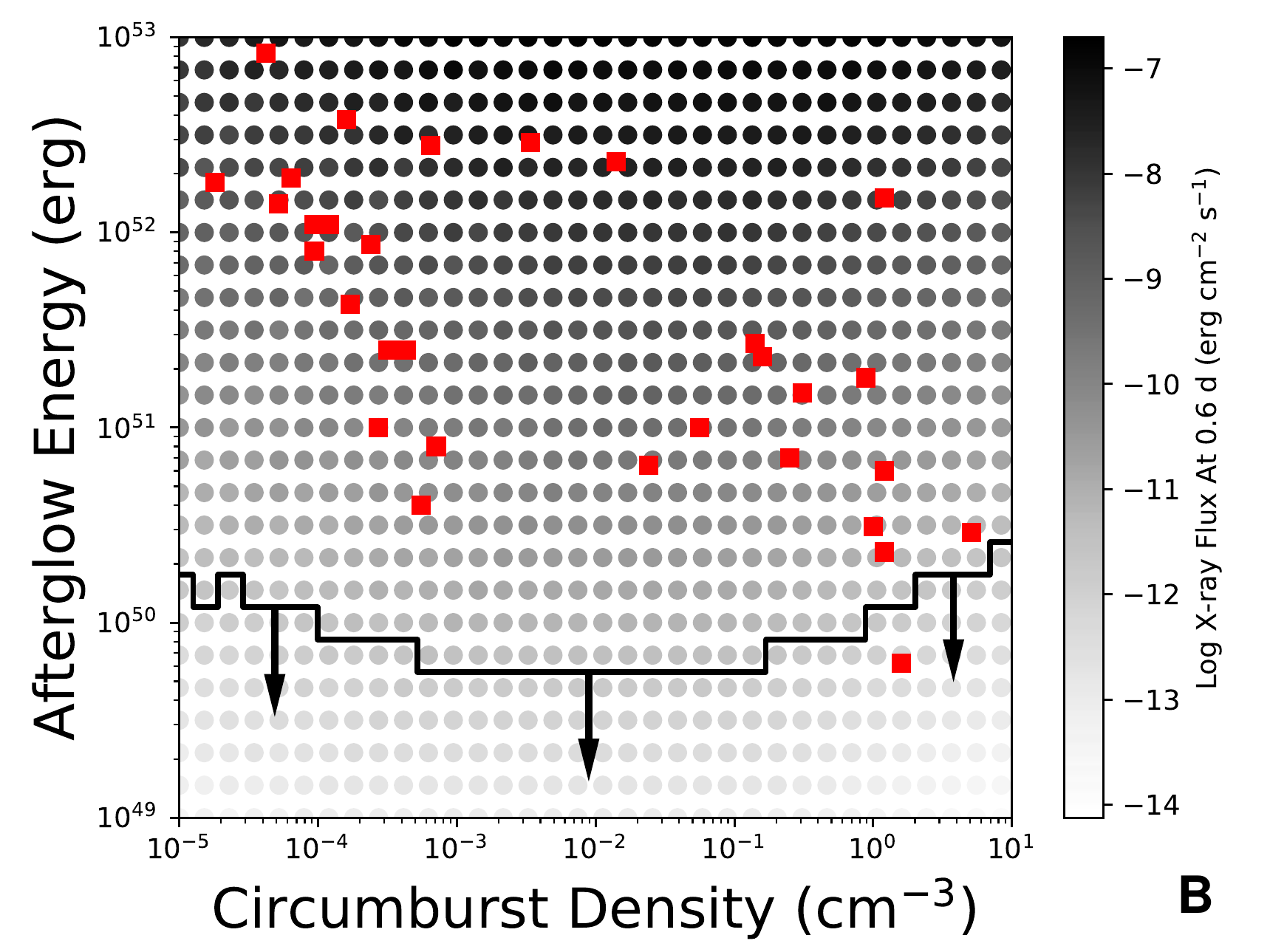}
\end{center}
\caption{{\bf Predicted X-ray flux of an afterglow to GW 170817.}
{\it A:} The distribution of short GRB light curves \cite{Davanzo14}, scaled to 40 Mpc. The solid line 
shows the median behavior; the two dashed lines represent the 25 and 
75 percentiles. The blue line with the triangle corresponds to the time range covered by 
the large-scale tiling with {\it Swift}-XRT and shows the typical 
sensitivity achieved per tile. The red arrows represent the XRT upper 
limits on emission from EM\,170817 obtained by summing all the
data up to the time of the arrow. The grey diamonds show the {\it NuSTAR} 
limits on emission from EM\,170817. {\it B:} 
The X-ray flux predicted for an on-axis jet for a range of isotropic afterglow 
energies and circumburst densities.  The black line indicates the flux 
upper limit of the first \emph{NuSTAR} observation; red squares are known 
short GRBs with $E_{\mathrm{AG}}$ and $n_{0}$ \cite{Fong15}.  Our observations rule out an energetic, 
ultra-relativistic outflow with $E_{\mathrm{AG}} > \sim 10^{50}$\,erg 
for on-axis geometries.
}
\label{fig:xray}
\end{figure}

In the standard model of GRBs \cite{Meszaros93,Piran04}, the prompt $\gamma$-ray
emission is generated by internal processes in a highly collimated,
ultra-relativistic jet.  As the ejecta expand and shock heat the
circumburst medium, electrons are accelerated and emit a broadband
synchrotron afterglow.  Our  UV and X-ray
observations place strong constraints on the presence and/or 
orientation of such ejecta following GW\,170817.

In Figure~\ref{fig:xray} we plot the median and 25--75\% distribution of
short GRB afterglows \cite{Davanzo14}, scaled to the
distance of NGC~4993.  While a handful of 
short GRBs have extremely fast-fading afterglows \cite{Rowlinson10} 
which would have been missed by our observations, the bulk of the 
population would have been easily detectable \cite{E17SM}.

We can translate these X-ray upper limits to physical constraints using the 
standard analytic afterglow formulation for synchrotron emission \cite{E17SM}.
We find that for on-axis viewing geometries, our non-detections limit the amount of energy 
coupled to relativistic ejecta ($E_{\mathrm{AG}}$) to be $E_{\mathrm{AG}} <\sim
10^{50}$\,erg (assuming the energy is radiated isotropically). To verify this result, we ran a series of simulations 
using the afterglow light curve code \texttt{boxfit} \cite{vanEerten12}.  
Over the range of circumburst densities and afterglow energies inferred 
for short GRBs \cite{Fong15}, we calculated the X-ray flux at the time of 
our first \emph{NuSTAR} epoch (which provides the tightest constraints, given typical afterglow
decay rates).  The results are shown in Figure~\ref{fig:xray}, yielding a 
similar constraint ($<\sim 10^{50}$\,erg) on the afterglow energy as our analytic approach.

Our X-ray upper limits also help to rule out an afterglow origin for the
UV emission: the optical to X-ray spectral index $\beta_{OX} \geq 1.6$
at $\Delta t = 0.6$\,d, is highly inconsistent with observed GRB
afterglows \cite{vkg+09}.  Analysis of the UV/optical spectral energy
distribution (SED) at early times ($\Delta t \leq 2$\,d) further supports
this conclusion \cite{E17SM}.  Fitting the SED with 
a blackbody function yields a temperature:
$T_{\mathrm{BB}}(\Delta t = 0.6$\,d$) = 7300 \pm 200$\,K, and
$T_{\mathrm{BB}}(\Delta t = 1.0$\,d$) = 6400 \pm 200$\,K (Figure~\ref{fig:uv}). A
power-law model, as would be expected for synchrotron afterglow
radiation, provides a very poor fit to the data \cite{E17SM}. We therefore conclude
that the observed UV counterpart must arise from a different physical
process than an on-axis GRB afterglow.

Given the apparent absence of energetic, ultra-relativistic material along
the line of sight, the detection of a short GRB is somewhat puzzling.
The isotropic gamma-ray energy release of GRB\,170817A, $E_{\gamma,\mathrm{iso}} = 
(3.08 \pm 0.72) \times 10^{46}$\,erg, is several orders of magnitude below any known 
short GRB \cite{LVCGBM}.  But even using the 
observed correlation \cite{Davanzo14} between $E_{\gamma,\mathrm{iso}}$ 
and X-ray afterglow luminosity, the predicted X-ray flux at $\Delta t = 
0.6$\,d is still above our \emph{Swift} and \emph{NuSTAR} upper limits.

This requires an alternative explanation for the observed $\gamma$-ray emission,
such as a (typical) short GRB viewed (slightly) off-axis,
or the emission from a cocoon formed by the interaction of a jet with the
merger ejecta \cite{Gottlieb17,Lazzati17,Kathirgamaraju17}.  We return to
this issue below in the context of late-time ($\Delta t > \sim 10$\,d)
X-ray emission (see also \cite{Kasliwal17} and \cite{Hallinan17}).

\section*{Implications of the Early UV Emission}

While inconsistent with ultra-relativistic ejecta (e.g.\ a GRB afterglow),
our UVOT observations nonetheless imply an ejecta velocity that is a 
substantial fraction of the speed of light.  If we convert the effective
radii derived in our SED fits (Figure~\ref{fig:uv}) to average velocities,
$\bar{v} \equiv R_{\mathrm{BB}} / \Delta t$ ($R_{\mathrm{BB}}$ is the radius of the emitting photosphere, $\Delta t$ is the
time delay between the trigger and the SED), we find that
$\bar{v}(\Delta t = 0.6$\,d$) \approx 0.3c$, and 
$\bar{v}(\Delta t = 1.0$\,d$) \approx 0.2c$ 
\cite{LVCC21577,LVCC21578}.  These velocities are much larger
than seen in even the fastest known supernova explosions \cite{Modjaz16}.
Similarly, the rapid cooling of the ejecta, resulting in extremely red
colors at $\Delta t \geq 1$\,d (Figure~\ref{fig:uv}), is unlike the 
evolution of any common class of extragalactic transient
\cite{Cowperthwaite15}.

Both of these properties are broadly consistent with theoretical predictions
for electromagnetic counterparts to binary neutron star mergers
known as kilonovae (sometimes called macronovae or mini supernovae)
\cite{Li98,Metzger17}.  
Numerical simulations of binary neutron star mergers
imply that these systems can eject $\sim 10^{-3}$--$10^{-2}$ solar masses
($M_{\odot}$) of material with $v \sim 0.1$--$0.2c$, either via tidal stripping and
hydrodynamics at the moment of contact (hereafter referred to as 
dynamical ejecta \cite{Rosswog99}),
or by a variety of processes after the merger, which include viscous, magnetic
or neutrino-driven outflows from a hyper-massive neutron star (if this is at least the temporary post-merger remnant)  and
accretion disc \cite{metzger09,perego14,martin15,Just15}. All of
these post-merger outflows are expected to have a less neutron rich composition
than the dynamical ejecta and in this study we use the
general term winds to refer to them collectively.

Next we examine the implications of the relatively
bright UV emission at early times.  Such UV emission
is not a generic prediction of all kilonova models: large opacity in the 
ejecta due to numerous atomic transitions of lanthanide elements can suppress UV emission, even at early times \cite{kbb13,bk13}.
This is particularly true for the dynamical ejecta, where a large 
fraction of the matter is thought to be neutron rich ($Y_e\leq 0.2$) and so produces
high atomic number elements (with $\sim$126 neutrons) via rapid neutron capture (the $r$-process,   \cite{Burbidge57}).

In contrast to the dynamical ejecta, a wind can have a significantly
larger electron fraction, particularly if irradiated by neutrinos.  $Y_{e}$
values of $\sim 0.2$ have been inferred from accretion discs around rapidly
spinning black holes \cite{Fernandez15b}, while a long-lived hyper-massive neutron star
may increase the neutrino flux even further ($Y_{e} \sim 0.3$; \cite{perego14}).
As a result of these large electron fractions, nucleosynthesis is expected to stop at
the second or even first $r$-process peak (elements with 82 or 50 neutrons respectively), resulting in few (if any) lanthanide
elements and a dramatically reduced opacity.

Our X-ray non-detections place limits on the presence of a long-lived
hyper-massive neutron star \cite{E17SM}.  In particular,
we can rule out any plausible neutron star remnant with a strong magnetic field that lived past the 
time of our first \emph{Swift} and \emph{NuSTAR} observations (which would effectively be 
a stable remnant, given the viscous time scale of the accretion 
disc).  Nonetheless, a short-lived or low-magnetic field hyper-massive neutron star, or a rapidly 
spinning black hole would both be consistent with our results. 

To investigate the plausibility of a  wind origin for the early UV 
emission, we have produced a series of 2-D models, 
varying the ejecta properties (mass, velocity, composition; \cite{wkf+17,E17SM}). We assume that the tidal
ejecta are more neutron rich ($Y_e\approx0.04$) than the 
wind ejecta ($Y_e\approx$0.27--0.37), and produce a sizable 
fraction of lanthanides that obscure the optical and UV emission. The spatial 
distribution 
of this high-opacity ejecta is based on merger models \cite{rosswog14a}. Obscuration 
by the disc formed from this high-opacity material causes a viewing-angle 
effect \cite{wkf+17}. 

In order to reproduce the early UV emission, we require models
with a wind ejecta mass $> \sim 0.03$\,$M_{\odot}$.  Furthermore,
a modest electron fraction ($Y_{e} \approx 0.27$), with significant 
amounts of elements from the first $r$-process peak, is strongly favored
over larger $Y_{e}$ ejecta ($Y_{e} \approx 0.37$, corresponding to mostly
Fe-peak elements).

The presence of bright UV emission strongly constrains
the observer viewing angle of the binary neutron star merger.  Sight lines
in the plane of the merger are expected to exhibit dramatically reduced UV 
emission due to the presence of the Lanthanide-rich dynamical ejecta.
For a wind mass ($M_{\mathrm{wind}}$) $\approx 0.03$\,$M_{\odot}$, a viewing angle of 
$\theta_{\mathrm{obs}} < \sim 30^{\circ}$ with respect to the rotation axis
is preferred.  Orientations up to $\sim 40^{\circ}$
can be accommodated with $M_{\mathrm{wind}} \approx 0.1$\,$M_{\odot}$;
at larger viewing angles the wind ejecta mass becomes unphysically
large.

While the wind component can provide a good fit to the UV emission,
on its own it under-predicts the observed optical/near-infrared flux at this
time.  Adding dynamical ejecta with $M_{\mathrm{dyn}} \approx 
0.01$\,$M_{\odot}$ and $v \approx 0.3c$ can provide a reasonable fit to
the early SEDs (Figure~\ref{fig:uv}).  However, we emphasize that
the properties of the dynamical ejecta are only poorly constrained at
early times; analysis of the full optical/near-infrared
light curve is necessary for accurate constraints on the Lanthanide-rich
material (e.g.\ \cite{Kasliwal17}). 


While much of the gamma-ray emission generated during the $r$-process
is re-radiated at optical/near-infrared wavelengths, it may also be
possible to observe directly emission lines from $\beta$-decay in the
{\it NuSTAR} bandpass.  We have calculated the expected signal from 10--100\,keV 
for a range of ejecta masses, and it is well below the {\it NuSTAR} limits
for GW\,170817 \cite{E17SM}.

The above modeling of the kilonova emission assumes that the merger ejecta
is unaffected by any energetic jet (or that no such jet is formed).  For jets
with a narrow opening angle ($\theta_{\mathrm{jet}} <~ 10^{\circ}$), numerical
simulations \cite{Gottlieb17} have shown that any such jet-ejecta interaction
will have negligible effects on the observed light curves on the time scales
probed by our observations.  

However, if the jet opening angle were sufficiently large, the energy from 
this jet (and the resulting cocoon) may accelerate material in the merger
ejecta to mildly relativistic velocities.  Numerical simulations in our companion paper 
\cite{Kasliwal17} offer some support for this scenario, providing a 
reasonable fit to the temperature and bolometric luminosity evolution
of EM\,170817.  However, they lack the detailed radiation transport calculations
presented here.

\section*{Late-Time X-ray Emission: Off-Axis Jet or Cocoon}

While no X-ray emission at the location of EM\,170817 was detected by
\emph{Swift} or \emph{NuSTAR}, a faint 
X-ray source was detected by {\it Chandra} at $\Delta t \approx 9$\,d  \cite{LVCC21765},
although the flux was not reported. Subsequent {\it Chandra} observations
at $\Delta t \approx 15$\,d reported $L_{X} \approx 9 \times 10^{38}$\,erg\,s$^{-1}$
\cite{Troja17,Haggard17}.  A variety of models predict long-lived
X-ray emission at a level $> \sim 10^{40}$\,erg\,s$^{-1}$ following
the merger of two neutron stars.  For example, (quasi-)isotropic X-ray
emission may be expected due to prolonged accretion onto a black hole
remnant, or from the spin-down power of a long-lived hyper-massive
neutron star. These models are not consistent with the {\it Swift} or 
{\it NuSTAR} limits, or the {\it Chandra} flux \cite{E17SM}, suggesting 
that, if a magnetar formed after the merger event, it collapsed to a black 
hole before our first X-ray observation (i.e. within 0.6 d of formation).
 
A possible explanation for the late-time X-ray emission is an 
off-axis (orphan) afterglow \cite{Rhoads99}.  If the binary neutron star
merger produces a collimated, ultra-relativistic jet, initially no emission
will be visible to observers outside the jet opening angle.  
As the outflow decelerates, the relativistic beaming becomes weaker and the
jet spreads laterally, illuminating an increasing fraction of the sky.
Off-axis observers can expect to see rising emission until the
full extent of the jet is visible, at which point the decay will appear
similar to that measured by on-axis observers. Simulations of such events
showed that starting a few days after the merger, off-axis afterglows 
represent the dominant population of GW counterparts 
detectable by {\it Swift} \cite{Evans16}. 

\begin{figure}
\begin{center}
\includegraphics[width=12cm,]{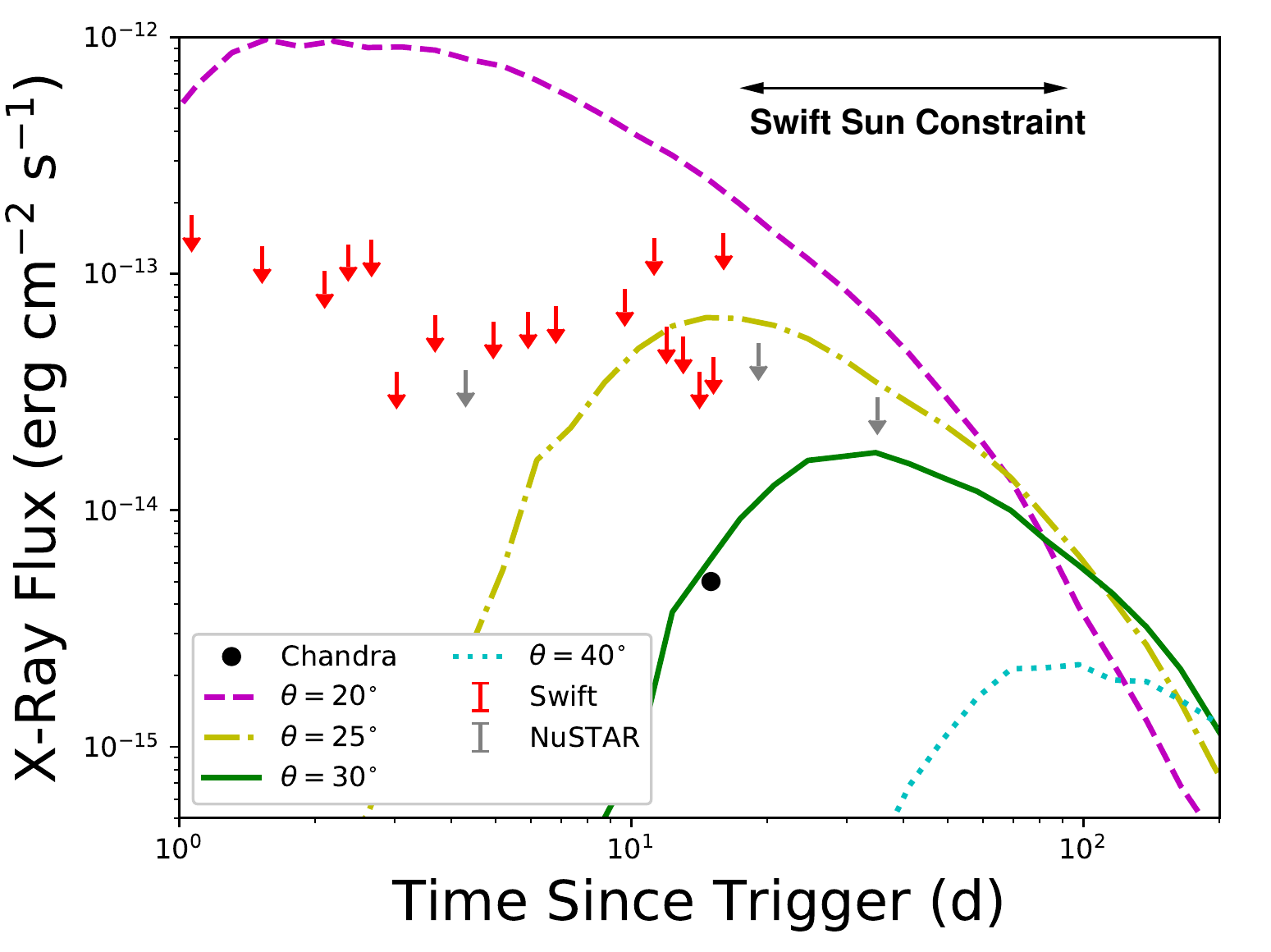}
\end{center}
\caption{{\bf Simulated X-ray afterglow light curves for typical short GRB 
parameters}  \cite{vanEerten12} . Here, $E_{\mathrm{AG}} = 2 \times 10^{51}$\,erg, 
$n_{0} = 5 \times 10^{-3}$\,cm$^{-3}$), and $\theta_{\mathrm{jet}} = 0.2$\,radian;
the true values of
these parameters are uncertain and vary between GRBs. 
Curves are shown for a range of viewing angles, with 
the \emph{Swift}-XRT and \emph{NuSTAR} limits marked. 
An off-axis orientation of $\approx 30^{\circ}$ is consistent with both the early \emph{Swift}-XRT and \emph{NuSTAR} 
limits, and the recently reported \textit{Chandra} detection 
\cite{LVCC21765}.  The anticipated peak time will occur when \emph{Swift}
and \emph{Chandra} cannot observe the field due to proximity to the Sun.}
\label{fig:offaxis}
\end{figure}

We ran a series of simulations using the \texttt{boxfit} code 
\cite{vanEerten12} to utilize our X-ray limits and the reported \emph{Chandra}
detections to constrain the orientation of GW\,170817 \cite{E17SM}.  
For the median values of short GRB afterglow energy, $E_{\mathrm{AG}} = 2
\times 10^{51}$\,erg, circumburst density, $n_{0} = 5 \times 
10^{-3}$\,cm$^{-3}$, and jet opening angle, $\theta_{\mathrm{jet}} = 0.2$ radian (12$^\circ$; \cite{Fong15}), the resulting light curves are 
plotted in Fig.~\ref{fig:offaxis}.  With the \emph{Swift} and \emph{NuSTAR} 
non-detections, these models rule out any viewing angle with
$\theta_{\mathrm{obs}} <\sim 20^{\circ}$.  Assuming the emission reported
by \textit{Chandra} results from an orphan afterglow, we infer 
$\theta_{\mathrm{obs}} \approx 30^{\circ}$.  

This inferred orientation is entirely consistent with the results of 
our analysis of the early UV emission.  However,
it is difficult to simultaneously explain the observed gamma-ray 
emission in this scenario, as it would require a viewing angle only
slightly outside the jet edge \cite{Kasliwal17}.  Either the observed
GRB\,170817A is powered by a source distinct from this jet, or we
are forced to disfavor an orphan afterglow model for the late-time
X-ray emission.

\begin{figure}
\begin{center}
\includegraphics[width=12cm,]{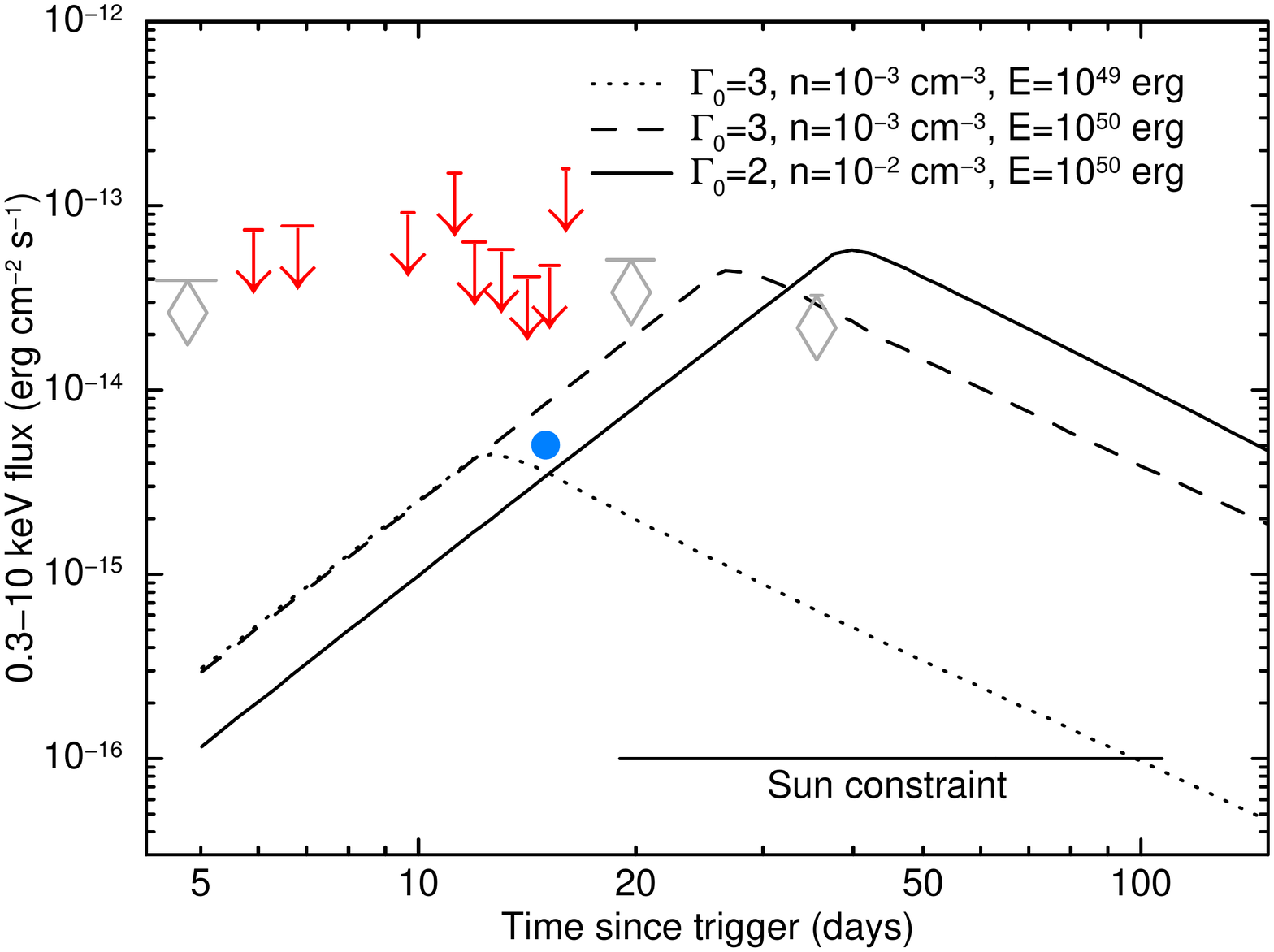}
\end{center}
\caption{{\bf Predicted X-ray light curves from a mildly relativistic jet}. The jet is based on model predictions
\cite{Gottlieb17}, for a range of different values
for the inital bulk Lorentz factor of the cocoon ($\Gamma_0$), circumburst density
($n$) and cocoon energy ($E$). Data points are the {\it Swift}-XRT (red arrows)
and {\it NuSTAR} (grey diamonds) upper limits and the {\emph Chandra} detection (blue) of EM\,170817.
The range of plausible peak times is not observable by \emph{Swift} (or \emph{Chandra}).}
\label{fig:cocoon}
\end{figure}

Alternatively, delayed X-ray emission may result if the initial outflow
speed is mildly relativistic, as would be expected from models where 
propagation in the merger ejecta forms a hot cocoon around the jet
\cite{Gottlieb17}.  In this case the rise is dictated by the time necessary for
the cocoon to sweep up enough material in the circumburst medium to 
radiate efficiently; this in turn depends on the energy carried by the expanding cocoon,
its bulk Lorentz factor and the circumburst density. Figure~\ref{fig:cocoon}
shows X-ray light curves predicted by this model for a range of
plausible values of these parameters, along with the X-ray limits 
from {\it NuSTAR} and {\it Swift}-XRT and the {\it Chandra} detection 
\cite{Troja17,Haggard17}. The latest {\it NuSTAR} datapoint disfavors energetic
cocoon models, particularly those at high density.  But lower energy or 
density models can fit all the X-ray data, whilst simultaneously accounting
for the gamma-ray emission \cite{Kasliwal17}. 

Our inferences regarding the origin of the late-time X-ray emission are 
broadly consistent with the conclusions reached in our companion radio paper
\cite{Hallinan17}.  Both an orphan afterglow and a mildly relativistic cocoon 
model make specific predictions for the evolution of the broadband flux over 
the upcoming months after the merger (Figures~\ref{fig:offaxis}-\ref{fig:cocoon};
\cite{Hallinan17}).

\section*{Conclusions}

The discovery of a short GRB simultaneous with a GW binary neutron star
merger represents the start of a new era of multi-messenger astronomy. 
It confirms that binary neutron star mergers can generate short gamma-ray 
transients \cite{Eichler89}, though the connection to classical short GRBs remains
unclear. Furthermore, GW\,170817 provides robust evidence that 
$r$-process nucleosynthesis occurs in the 
aftermath of a binary neutron star merger \cite{Kasliwal17}.  

While a kilonova detection following a short GRB has been previously reported \cite{Tanvir13a,Berger13}, our
multi-wavelength dataset has allowed us confront kilonova models with 
UV and X-ray observations.  
The absence of X-ray emission largely rules out the presence of an
energetic, ultra-relativistic, and collimated outflow viewed from within 
the opening angle of the jet.  The  
late-time X-ray emission is consistent with a collimated, 
ultra-relativistic outflow viewed at an off-axis angle of $\approx30^{\circ}$
(i.e.\ an orphan afterglow).  A mildly relativistic outflow,
as may be expected if the jet were enveloped by a hot cocoon, 
is also consistent with our X-ray data (and may naturally explain the 
peculiar properties of the gamma-ray emission; \cite{Kasliwal17}).

The presence of bright, rapidly fading UV emission was not a generic 
prediction of kilonova models and requires special circumstances to avoid
obscuration by the heavy elements formed in the dynamical ejecta.
We find that we can reproduce the early UV and optical emission with a
massive ($M \approx 0.03 M_{\odot}$) and high-velocity ($v \approx 
0.08c$) outflow comprised of moderate-$Y_{e}$ (first $r$-process peak)
material at a viewing angle of $\approx 30^{\circ}$; such winds
may be expected if the remnant is a relatively long-lived hyper-massive
neutron star or a rapidly spinning black hole.  Alternatively, if the 
hot cocoon is able to accelerate 
material in the ejecta to mildly relativistic speeds, this may also
be able to account for the early UV emission \cite{Kasliwal17}.

\section*{Acknowledgments}

This paper is dedicated to the memory of Neil Gehrels 
(1952--2017), who was PI of {\it Swift} until his untimely death
in February 2017. Neil's inspirational leadership was the lifeblood of {\it Swift} and he was always keen to innovate
and expand the mission's capabilities. Without his direction, leadership and support, the work we present here would not have been possible.

Funding for the {\it Swift} mission in the UK is provided by the UK Space Agency. SRO
gratefully acknowledges the support of the Leverhulme Trust Early Career
Fellowship (SRO). The {\it Swift} team at the MOC at Penn State acknowledges support
from NASA contract NAS5-00136. The Italian {\it Swift} team acknowledge support from
ASI-INAF grant I/004/11/3. SR has been supported by the Swedish Research Council
(VR) under grant number 2016- 03657\_3, by the Swedish National Space Board
under grant number Dnr. 107/16 and by the research environment grant 
``Gravitational Radiation and Electromagnetic Astrophysical Transients (GREAT)''
funded by the Swedish Research council (VR) under Dnr 2016- 06012. This research
used resources provided by the Los Alamos National Laboratory Institutional
Computing Program, which is supported by the U.S. Department of Energy National
Nuclear Security Administration under Contract No. DE-AC52-06NA25396.
VLT data were obtained under ESO programme number 099.D-0668.
NuSTAR acknowledges funding from NASA Contract No. NNG08FD60C.

The observations are archived at http://www.swift.ac.uk for {\it Swift} and \\
https://heasarc.gsfc.nasa.gov/docs/nustar/nustar\_archive.html for {\it NuSTAR}
under the observation IDs given in Table S2. Reduced photometry and surveyed areas are
tabulated in the supplementary material. The {\sc boxfit} software is available at \\
http://cosmo.nyu.edu/afterglowlibrary/boxfit2011.html, {\sc SuperNu} at  \\
https://bitbucket.org/drrossum/supernu/wiki/Home,
Access to {\sc WinNet} source code and input files will be granted
upon request via: https://bitbucket.org/korobkin/winnet.
The dynamical model ejecta are available via
http://compact-merger.astro.su.se/downloads\_fluid\_trajectories.html (run 12)
The {\sc SuperNu} and {\sc boxfix} input files are available as Supplementary Files
at Science Online.

\clearpage
\beginsupplement

\begin{center}
\noindent
\Huge{Supplementary material for {\it Swift} and {\it NuSTAR} observations of GW170817: detection of 
a blue kilonova} 
\end{center}

\section*{Materials and Methods}

\section{Observations}

\begin{figure}
\begin{center}
\includegraphics[width=12.1cm]{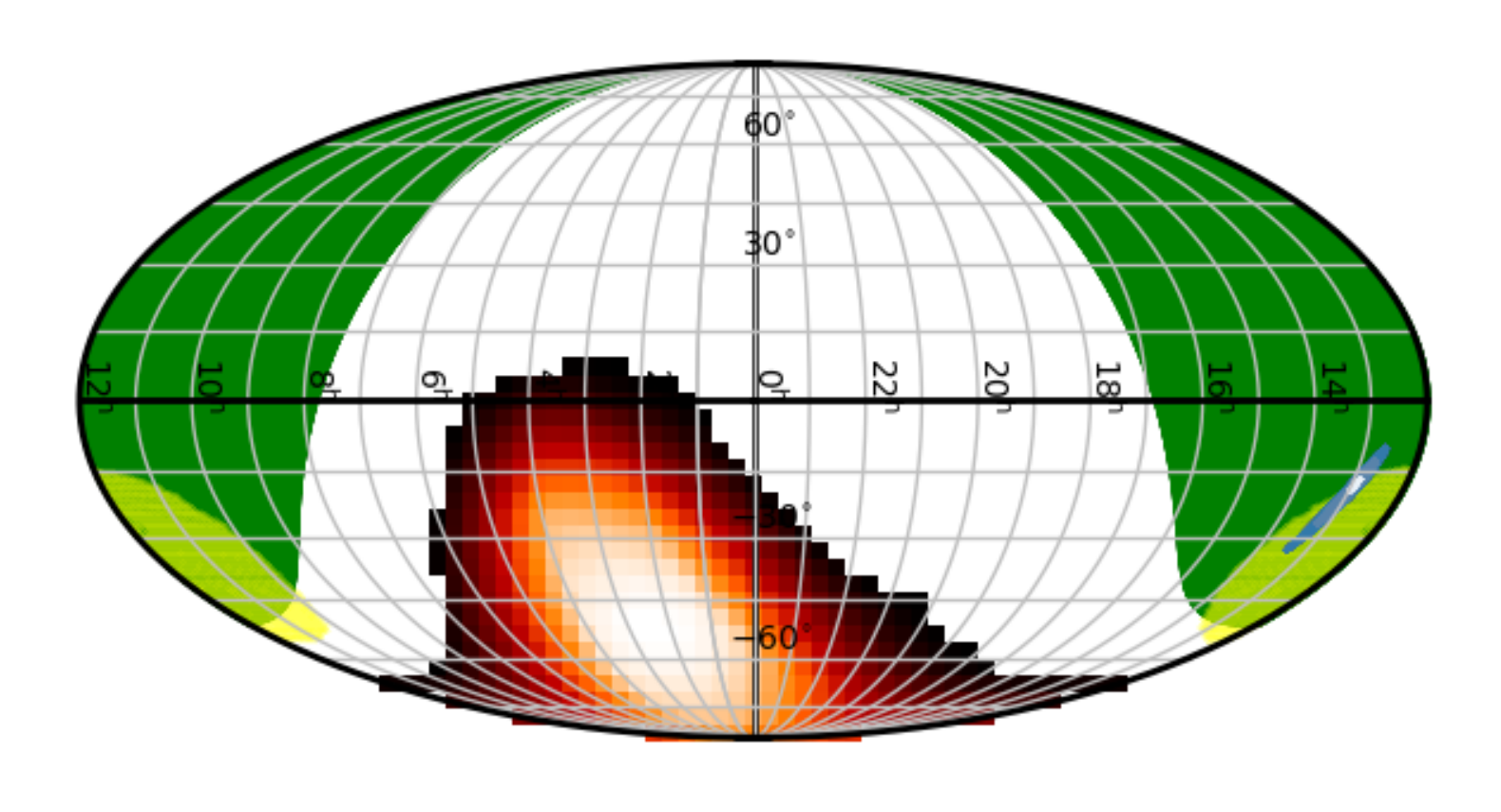}
\end{center}
\caption{{\bf The BAT field of view at the time of the GW/GRB trigger}. The black/red/white scale indicates the BAT coding fraction; the darker
colour indicates lower partial coding fraction (i.e.\ more toward the edge of the field of view).
The GW localizations probability contours are plotted in blue. The green region shows the part of the sky occulted by the Earth
and the yellow region is the {\it Fermi}-GBM 90\%\ confidence region.}
\label{fig:bat_fov}
\end{figure}

Advanced LIGO and advanced Virgo (operating as the LIGO/Virgo Consortium: hereafter `LVC') registered GW\,170817 at
12:41:04.45 UT on 2017 August 17. This was announced to the followup community first as an LVC/Gamma-ray Coordinates Network (GCN) notice at 13:08 UT,
and then by \cite{LVCC21505}, (trigger `G298048'). \emph{Fermi} Gamma-ray Burst Monitor (GBM) triggered on a weak short gamma-ray signal at
12:41:06 UT \cite{LVCC21506}, reported as a short GRB (sGRB, \cite{GCN21520}). A weak signal was also observed by the
INTEGRAL satellite \cite{LVCC21507}. At this time, the {\it Swift}\ Burst Alert Telescope (BAT; \cite{BarthelmyBAT}) was
pointing at RA, Dec (J2000)= 02$^h$24$^m$18.0$^s$, -52$^\circ$17$^\prime$13$^{\prime\prime}$, i.e.\ away from the GBM
and GW position. Indeed, the GW localization region was entirely occulted by the Earth (Figure~\ref{fig:bat_fov}), with none
of the probability region above the Earth limb (defined as 69$^\circ$ from the center of the Earth, which
accounts for the radius of the Earth plus 100 km of atmosphere, up to the K\'{a}rm\'{a}n line, which commonly represents
the top of the atmosphere). The position of EM\,170817A (below) was likewise not visible to {\it Swift} at the trigger
time, while the very edge of the GBM error region overlapped the edge of the BAT field of view (0.1\% of the GBM probability
was within the BAT field of view). We found
no evidence for a signal in the BAT within 100-s of the GW and GBM triggers.

{\it Swift}\ observations comprising more than one pointing can only be uploaded to the spacecraft in the form of a pre-planned science timeline
which requires a ground-station pass. In order to obtain some rapid coverage, we uploaded commands using the NASA
Tracking and Data Relay Satellite System (TDRSS) network to utilize the on-board 37-point tiling pattern. Using this
pattern the X-ray Telescope (XRT; \cite{BurrowsXRT}) covered a region $\sim$1.1$^\circ$ in radius, which was centered on
the GBM position. The UV/optical Telescope (UVOT; \cite{RomingUVOT}) has a smaller field of view
($17^\prime\times17^\prime$) than the XRT ($24.6^\prime$ diameter), thus there are gaps between the tiles in UVOT. These
observations began at 13:37 UT, 0.04 d after GW\,170817.

The initial GW skymap was created using only a single GW detector (LIGO-Hanford; \cite{LVCC21509}) and thus covered almost
the entire sky. The GBM error had a statistical position error of 11.6$^\circ$ radius (1-$\sigma$) with a systematic
error estimated to be at least 3.7$^\circ$ \cite{LVCC21506}; the GBM localization probability has a 90\%\ 
confidence area of 1626 deg$^2$. We convolved the GW skymap with the GWGC catalog
\cite{White11}, using the 3D information in the GW skymap and the methodology described by \cite{Evans16}. This
effectively re-weights the GW probability according to our knowledge of the location of the stellar mass in the local
Universe, accounting for our estimate of how incomplete that knowledge is. We next multiplied the resultant probability
map by a map of the GBM localization probability and renormalised the result, then identified the most probable XRT
fields-of-view from this skymap. {\it Swift}-XRT and UVOT observations of these fields, with a nominal 120 s per
exposure began at 17:16 UT, 0.19 d after GW\,170817. The onboard slewing algorithm is not perfectly modelled, thus exposure times vary someone
from the nominal value. 

At $\sim$17:30 UT a revised GW skymap was released which used data from all three LVC interferometers and had a 90\%\
credible region of 31\,deg$^{2}$ \cite{LVCC21513}. Since this error region only overlapped the outer contours of the
GBM error region, our planned observations did not significantly cover this revised localisation. We therefore convolved
this new skymap with GWGC as above, and generated a new observing plan based on this probability map; these observations
began at 20:24 UT, 0.31 d after GW\,170817, and are listed in Table~\ref{tab:obs}.

At 01:05 UT on 2017 August 18, \cite{LVCC21529} reported the detection of a candidate EM counterpart, hereafter denoted
`EM\,170817', apparently associated with the galaxy NGC~4993 (distance$\approx$40~Mpc \cite{Freedman01}). Shortly
thereafter, several others groups reported detection of the source \cite{LVCC21530,LVCC21531}, sometimes with differing
names (e.g.\ SSS17a, DLT17ck). For simplicity we refer to the object henceforth as EM\,170817.

{\it Swift}-XRT and UVOT observed this field for 2 ks, starting on 2017 Aug 18 at 03:35 UT ($\Delta t = 0.6$\,d),
temporarily overriding the wide-area search of the GW error region. The optical source was detected in all UV filters,
but no X-ray counterpart was seen \cite{LVCC21550}.

The nature of EM\,170817 was not clear at this time, so {\it Swift} resumed the wide-area search of the GW error region,
but regularly interrupted it to carry out repeat observations of EM\,170817. These targeted observations continued after
the large-area tiling was complete (at $\Delta t=4.3$ d), apart from the interval 22:04 on August 24 to 22:14 on August
26, at which times the source was too close to the Moon for {\it Swift} observations.

After the discovery of X-ray emission from EM\,170817 by {\it Chandra} \cite{LVCC21765,Troja17,Haggard17} we increased the frequency of
these observations, until the field entered the {\it Swift} Sun constraint at 05:46 on September 2, from which it will
emerge on December 02. Details of the {\it Swift} observations of EM\,170817 are given in Table~\ref{tab:xray}.

The \emph{Nuclear Spectroscopic Telescope ARray} ({\it NuSTAR}) \cite{Harrison13} also observed the location of
EM\,170817, taking data at four different epochs starting at 4.8 d after the GW trigger. Details of these observations
are also presented in Table~\ref{tab:xray}.

\section{Data analysis}

\subsection{{\it Swift}-BAT}

While the location of EM\, 170817A was not visible to {\it Swift} at the time of the trigger,
the initial GW localisation (which covered almost the entire sky) and the edge of the {\it Fermi}
localization error were within the BAT field of view. We therefore
searched the raw 15--350 keV BAT light curves for any evidence for GRB emission within 100-s of the GW/GBM trigger. No such emission was detected.
As there was no BAT trigger, this search is limited to the raw 15--350 keV light curves (the so-called `rate data' accumulated onboard), which have
time bins of 64ms, 1 s and 1.6 s. Assuming a sGRB with a typical spectrum in the BAT energy range (i.e.\ a simple power-law model with a power-law index
of $-1.32$, from the averaged sGRB power-law indices based on \cite{Lien16}), the 5-$\sigma$ upper limit in the 1.6-s
binned light curve corresponds to a 15--350 keV flux of $<2.3 \times 10^{-8}$ erg cm$^{-2}$ s$^{-1}$ for an
on-axis  energy response of the BAT, and $<4.8 \times 10^{-7}$ erg cm$^{-2}$ s$^{-1}$ for an
extremely off-axis (i.e\, within the {\it Fermi} error region) energy response of the BAT.

\textit{Swift} began observations of EM\,170817 with the UVOT at 03:37 on
2017 August 18 ($\Delta t = 0.6$\,d), initially with just the  
$uvw2$, $uvm2$, $uvw1$, and $u$ filters.  Observations with the full 
6-filter complement (including the $v$ and $b$ bands) began at 
$\Delta t = 2.5$\,d using a pointing offset (required to avoid to a nearby bright 
star). The source was detected in all UV filters; $b$ and $v$ band yield
only marginal detections or upper limits, in part due to the considerable 
galaxy background. UVOT analysis was carried out using  {\sc heasoft}
v6.21 \cite{heasoft} and UVOT CALDB version 20170130 \cite{caldb}.

Absolute photometry was obtained by extracting the photon counts 
using a circular aperture of 3\arcsec\ radius centered on EM\,170817, followed 
by an aperture correction to the calibration standard of 5\arcsec. The use of this smaller
aperture reduced the contamination from NGC\,4993 and increased the signal-to-noise ratio 
\cite{Poole08}.  Background counts were extracted from a region located well 
outside of the host galaxy. Count rates (including coincidence loss correction) 
were obtained using the {\sc heasoft} tool {\sc uvotsource}.

To determine the host galaxy contribution to the source, we measured the 
background-subtracted count rate in the source region in the interval $\Delta 
t = 9.5$--$16$\,d, by which time the source had faded below detectability in 
all filters. A few images were removed due to an unstable spacecraft attitude. The 
host count rate was then subtracted from the earlier measurements of EM\,170817; 
this was performed for each filter and exposure individually. The 
host-subtracted count rates were then converted to AB magnitudes using zero 
points from \cite{Breeveld11}.

Examination of the spectral energy distribution (SED) derived from the photometry 
showed that the flux drops off steeply going further into the UV. As a 
result the effective wavelength of the UV bands is shifted to the red. We 
calculated the effective wavelengths of the UVOT bands by iteratively 
computing the weighted average of the wavelength multiplied by the effective area 
multiplied by the approximate flux (as derived from the SED), normalized by the effective area and flux integral. 
The results are shown in Table~\ref{tab:uvot}.

Finally, light curves were corrected for Galactic extinction, $E(B-V) = 
0.106$ \cite{sf11}, derived using a region of size 2$^{\circ}$. The final, 
background-subtracted light curves are shown in Figure 2A.

\begin{table}
\begin{tabular}{lllllllll}
Filter & $\lambda_{\mathrm{eff}}$ & $\Delta t$ & Duration & Count Rate
  & AB Magnitude \\
       & ($\AA$) & (d) & (d) & (s$^{-1}$) &  \\
\hline
\hline
uvw2 & 2296 & 0.6431 & 0.0029 & $0.151 \pm 0.032$ & $21.13_{-0.21}^{+0.26}$\\
uvw2 & 2366 & 1.0469 & 0.0029 & $0.041 \pm 0.024$ & $>21.45$ \\
uvw2 & $\cdots$ & 1.5067 & 0.003  & $0.026 \pm 0.022$ & $>21.66$ \\
uvw2 & $\cdots$ & 2.0992 & 0.0646 & $0.027 \pm 0.014$ & $>21.98$ \\
uvw2 & $\cdots$ & 2.3612 & 0.0702 & $0.021 \pm 0.013$ & $>22.16$ \\
uvw2 & $\cdots$ & 2.6601 & 0.0376 & $0.013 \pm 0.014$ & $>22.21$ \\
uvw2 & $\cdots$ & 3.0286 & 0.196  & $0.0195 \pm 0.0088$ & $>22.42$ \\ 
\hline
uvm2 & 2365 & 0.6272 & 0.0043 & $0.092 \pm 0.018$ & $21.12_{-0.20}^{+0.24}$ \\
uvm2 & 2405 & 1.0652 & 0.0308 & $0.026 \pm 0.011$ & $22.52_{-0.39}^{+0.62}$ \\
uvm2 & $\cdots$ & 1.5237 & 0.0378 & $0.0122 \pm 0.0087$ & $>22.07$ \\
uvm2 & $\cdots$ & 2.0723 & 0.0336 & $0.012 \pm 0.010$ & $>21.97$ \\
uvm2 & $\cdots$ & 4.9533 & 0.4697 & $0.0062 \pm 0.0068$ & $>22.47$ \\
\hline
uvw1 & 2590 & 0.6344 & 0.0029 & $0.641 \pm 0.064$ & $19.46_{-0.10}^{+0.11}$ \\
uvw1 & 2790 & 1.0407 & 0.0016 & $0.321 \pm 0.062$ & $20.21_{-0.19}^{+0.23}$ \\
uvw1 & $\cdots$ & 1.5293 & 0.0345 & $0.042 \pm 0.029$ & $>21.20$ \\
uvw1 & $\cdots$ & 2.0777 & 0.0350 & $0.004 \pm 0.023$ & $>21.79$ \\
uvw1 & $\cdots$ & 3.0211 & 0.1960 & $0.004 \pm 0.018$ & $>22.05$ \\
\hline
u & 3529 & 0.6387 & 0.0014 & $2.93 \pm 0.22$ & $18.19_{-0.08}^{+0.09}$ \\
u & 3666 & 1.0431 & 0.0008 & $1.39 \pm 0.20$ & $19.00_{-0.15}^{+0.17}$ \\
u & $\cdots$ & 1.5021 & 0.0015 & $0.27 \pm 0.11$ & $20.79_{-0.39}^{+0.61}$ \\
u & $\cdots$ & 2.3546 & 0.0674 & $0.098 \pm 0.094$ & $>20.41$ \\
u & $\cdots$ & 4.9461 & 0.4673 & $0.038 \pm 0.072$ & $>20.85$ \\
\hline
b & $\cdots$ & 2.3565 & 0.0674 & $0.27 \pm 0.16$ & $>19.31$ \\
b & $\cdots$ & 2.6554 & 0.0348 & $0.42 \pm 0.19$ & $19.93_{-0.40}^{+0.65}$ \\
b & $\cdots$ & 3.0258 & 0.1951 & $0.15 \pm 0.12$ & $>19.71 $ \\
b & $\cdots$ & 3.6517 & 0.1006 & $0.26 \pm 0.15$ & $>19.37 $ \\
b & $\cdots$ & 5.9098 & 0.1654 & $0.22 \pm 0.14$ & $>19.50 $ \\
\hline
v & $\cdots$ & 2.3668 & 0.0683 & $0.11 \pm 0.12$ & $>18.72$ \\
v & $\cdots$ & 2.6657 & 0.0357 & $0.12 \pm 0.12$ & $>18.67$ \\
v & $\cdots$ & 3.0029 & 0.1628 & $0.172 \pm 0.097$ & $>18.72$ \\
v & $\cdots$ & 3.6620 & 0.1016 & $0.07 \pm 0.10$ & $>18.95$ \\
v & $\cdots$ & 5.9161 & 0.1641 & $0.19 \pm 0.12$ & $>18.54$ \\
\hline
\end{tabular}
\caption{{\bf \emph{Swift}-UVOT photometry of EM\,170817}.  The reported 
magnitudes have been corrected for host contamination but not for
Galactic extinction.  Upper limits are quoted at the 3~$\sigma$ 
level.  Epochs lacking sufficient SED constraints 
(and thus lacking estimates of the effective wavelength) are 
indicated with ``$\cdots$''.}
\label{tab:uvot}
\end{table}

\subsection{{\it Swift}-XRT}

The data were analysed using {\sc heasoft v6.21} \cite{heasoft} and CALDB version 20170501 \cite{caldb}. The XRT data were grouped into ``analysis
blocks'' of overlapping fields, to a maximum size of $\sim$50$^\prime$\ in diameter. Source detection was carried out on
each of these blocks using the iterative method developed by \cite{Evans14}. Any sources found were ranked according to
their likelihood of being a transient related to the GW trigger. The method and ranking is described in detail by
\cite{Evans16b}. No sources above rank 3 (i.e.\ uncataloged, but below catalog limits) were found in our search of
the GW error region. 

We focus only on the data at the location of EM\,170817. No source was found in the first 2-ks
observation of this source\cite{LVCC21550}, with a 0.3--10 keV 3-$\sigma$ upper limit of 5.5$\times10^{-3}$ ct s$^{-1}$.
Assuming an energy conversion of 4$\times10^{-11}$ erg cm s$^{-1}$\ ct$^{-1}$, which is typical for GRB afterglows
\cite{Evans09}, this corresponds to a flux of $2.2\times10^{-13}$ erg cm$^{-2}$ s$^{-1}$. Given the distance to
NGC~4993 of $\sim$40 Mpc this equates to $L_{\rm X} < 4.2\times10^{40}$ erg s$^{-1}$.

No X-ray emission was detected near the position of EM\,170817 in the next four observations, when a standard GRB
afterglow would be at its brightest (see, Figure~4 and \cite{Evans16}, their figures 12--13).
However, we did detect a source $\sim$ 2 days after the trigger, once we had accumulated 17.5 ks of exposure time with
XRT \cite{LVCC21612}. The source was poorly
localized due to its faintness, but lay 11.3$^{\prime\prime}$\ from the optical counterpart position, and
8.1$^{\prime\prime}$ from NGC~4993, with a position uncertainty of 6.4$^{\prime\prime}$ (90\%\ confidence). Summing all of
the X-ray data on this source, and correcting the astrometry by aligning the X-ray sources in the field with objects
cataloged in 2MASS as described by \cite{Evans14} (and using only XRT sources detected with flags `Good' or
`Reasonable', i.e.\ unlikely to be spurious detections.), we obtain a final position of RA, Dec (J2000)= 13$^{\mathrm{h}}$09$^{\mathrm{m}}$47.65$^{\mathrm{s}}$,
-23$^\circ$23$^\prime$01.6$^{\prime\prime}$, with an uncertainty of 3.9$^{\prime\prime}$\ (radius, 90\% confidence). This is
10.2$^{\prime\prime}$\ from the position of EM\,170817, and 0.9$^{\prime\prime}$\ from the position of NGC~4993, suggesting that
the emission detected is from the host galaxy, rather than the electromagnetic counterpart. This interpretation was confirmed by
\emph{Chandra} observations \cite{LVCC21648,Troja17,Haggard17,Margutti17}.  Taken with the detection of radio emission from the galaxy
\cite{LVCC21548,LVCC21559,Hallinan17,Alexander17} this may indicate an active galactic nucleus in NGC~4993; however the X-ray emission appears somewhat diffuse \cite{LVCC21648} and is thus likely
made up of multiple X-ray emitters in that galaxy. 

Since this emission may
contaminate the location of EM\,170817, the predicted count-rate from the host galaxy (assuming point-like emission) was
taken into account when calculating upper limits.
Summing all of the XRT data (172 ks), we calculate a 3-$\sigma$ upper limit at the location
of EM\,170817 of $2.8\times10^{-4}$ ct s$^{-1}$. Using the same assumptions as above, this corresponds to $L_{\rm X} <
2.1\times10^{39}$ erg s$^{-1}$ (0.3--10 keV). The count rates or upper limit for the host NGC~4993 and EM\,170817
for each individual observation are given in Table~\ref{tab:xray}.

\begin{table*}
\begin{center}
\caption{{\bf X-ray pointed observations of EM\,170817 with {\it Swift}-XRT and
{\it NuSTAR}}. For XRT we give the measured luminosity or 3-$\sigma$ upper limit of NGC 4993
and 3-$\sigma$ limit at the location of EM\,170817. For {\it NuSTAR} we give only the latter.}
\label{tab:xray}
\begin{tabular}{ccccc}
\hline
ObsID & Start time$^a$ & Exposure  & log $L_{X,\rm NGC~4993}$   & log $L_{X,\rm EM\,170817}$        \\
 & (UTC) & (ks)  & 0.3--10 keV  & (0.)3--10 keV$^b$   \\
 & & & erg s$^{-1}$  & erg s$^{-1}$ \\
\hline
{\it Swift} \\
07012978001  &  Aug 18 at 03:34:33 (0.62) &   2.00 & $<40.56$& $<40.72$ \\
07012167001  &  Aug 18 at 12:11:49 (0.98) &   0.12 & $<41.88$& $<41.70$ \\
07012978002  &  Aug 18 at 13:29:43 (1.03) &   1.99 & $<40.65$& $<40.53$ \\
07012978003  &  Aug 19 at 00:18:22 (1.48) &   2.99 & $<40.58$& $<40.40$ \\
07012978004  &  Aug 19 at 13:24:05 (2.03) &   4.98 & $<40.22$& $<40.29$ \\
07012979001  &  Aug 19 at 18:30:52 (2.24) &   4.97 & $40.10^{+0.18}_{-0.22}$& $<40.41$ \\
07012979002  &  Aug 20 at 03:24:44 (2.61) &   4.98 & $39.98^{+0.20}_{-0.25}$& $<40.43$ \\
07012979003  &  Aug 20 at 08:28:05 (2.82) &   14.96 & $<40.16$& $<39.87$ \\
07012979004  &  Aug 21 at 01:43:44 (3.54) &   9.97 & $<40.20$& $<40.11$ \\
07012979005  &  Aug 22 at 00:05:57 (4.48) &   9.90 & $<40.18$& $<40.08$ \\
07012979006  &  Aug 23 at 06:22:57 (5.74) &   9.04 & $<40.21$& $<40.12$ \\
07012979007  &  Aug 23 at 23:59:57 (6.47) &   10.37 & $39.83^{+0.17}_{-0.21}$& $<40.14$ \\
07012979008  &  Aug 26 at 23:59:57 (9.47) &   7.30 & $39.80^{+0.21}_{-0.26}$& $<40.22$ \\
07012979009  &  Aug 28 at 10:46:17 (10.92) &   3.37 & $<40.35$& $<40.43$ \\
07012979010  &  Aug 29 at 01:04:57 (11.52) &   11.78 & $39.98^{+0.15}_{-0.17}$& $<40.06$ \\
08012979011/12  &  Aug 30 at 01:00:57 (12.51) &   15.81 & $39.64^{+0.18}_{-0.23}$& $<40.02$ \\
07012979015  &  Aug 31 at 02:27:52 (13.57) &   34.88 & $39.90^{+0.09}_{-0.10}$& $<39.87$ \\
08012979014/17  &  Sep 01 at 05:53:04 (14.72) &   25.62 & $39.63^{+0.16}_{-0.19}$& $<39.93$ \\
08012979016/19  &  Sep 02 at 08:40:56 (15.83) &   3.76 & $<40.68$& $<40.46$ \\
\hline
{\it NuSTAR} \\
90361003002 & Aug 18 at 05:25 (0.7) & 23.9 & --- & $<39.7$ \\
90361003004 & Aug 21 at 20:45 (4.3) & 40.3 & --- & $<39.6$ \\
90361003005 & Sep 04 at 17:56 (18.2)& 12.6 & --- & $<40.1$ \\
90361003006 & Sep 05 at 14:51 (19.1)& 19.3 & --- & $<39.9$ \\
90361003007 & Sep 06 at 17:56 (20.1)& 23.1 & --- & $<39.8$ \\
90361003009 & Sep 21 at 11:10 (34.9) & 54.1 & -- & $<39.5$ \\
\hline
\end{tabular}
\end{center}
\begin{flushleft}
$^a$ Observation year is 2017, times in parentheses are days since GW\,170817.
$^b$ XRT luminosities are 0.3--10 keV, {\it NuSTAR} are 3--10 keV.
\end{flushleft}
\end{table*} 

\subsection{{\it NuSTAR}}

The {\it NuSTAR} data were reduced using {\sc nustardas} version 06Dec16\_01.7.1 within {\sc heasoft} \cite{heasoft}, and CALDB version 20170614 \cite{caldb}. We used all standard
settings in {\sc nupipeline}, but included filtering modes for the South Atlantic Anomaly ({\sc saacalc=2,
saamode=optimized, tentacle=yes}), to reduce the solar low energy background. For extracting count rates, we used an
optimized extraction region of radius 20$^{\prime\prime}$, which allows for a possible 1-$\sigma$ positional offset of
8$^{\prime\prime}$ in the {\it NuSTAR} absolute astrometry \cite{Harrison13}, and ensures that the {\it NuSTAR} full-width
half-maximum point-spread function (PSF), which has a diameter of 20$^{\prime\prime}$, is included in the extraction
region. The region was centered on the location of EM\,170817 \cite{LVCC21529}. Spectra and responses were obtained for each
module and epoch using {\sc nuproducts} with default settings.

We combined the data from the two {\it NuSTAR} focal plane modules (FPM): `FPMA' and `FPMB', and for the first epoch we obtained count rates of
$7.8\times10^{-4}$ ct s$^{-1}$ (3--10 keV), $1.2\times10^{-3}$ ct s$^{-1}$ (10--39 keV) and $7.5\times10^{-4}$ ct
s$^{-1}$ (39--79 keV). We compared these count rates with predicted background rates obtained from {\sc nuskybkg}, which
are based on a model fitted to the background of the entire module \cite{Wik14}, and found that they agreed at the
1-$\sigma$ level, indicating that the neither EM\,170817 nor NGC~4993 were detected. We calculated the 3-$\sigma$ upper limits on
emission from EM\,170817 by assuming a power-law spectrum with a photon index of $-2$ and absorption 
Hydrogen column density $3\times10^{21}$\,cm$^{-2}$, as is typical for GRBs \cite{Evans09},
and assuming a distance of 40 Mpc. These are given in Table~\ref{tab:xray}. We analyzed the other epochs in the same way; the source was likewise undetected, with the limits
given in Table~\ref{tab:xray}. Combining the data from all epochs yields a limiting luminosity of $L=2.8\times10^{39}$ erg s$^{-1}$ (3--10 keV),
$1.2\times10^{40}$ erg s$^{-1}$ (10--39 keV), and $1.7\times10^{41}$ erg s$^{-1}$ (39--79 keV). The non-detection of NGC~4993 in the
$>3$ keV band pass of {\it NuSTAR} suggests a relatively soft spectrum, given the detection by {\it Swift}-XRT. Indeed, we find no
evidence for an elevated count-rate at the location of NGC~4993. Thus, while we
cannot rule out that our {\it NuSTAR} limits on the flux from EM\,170817 are slightly underestimated due to contamination
by host-galaxy emission, any such contamination is small and will have little effect. 

\section{Interpretation and discussion}

\subsection{Gamma-Ray Emission}

If the GRB\,170817 was associated with GW\,170817, then it is the
closest sGRB with known distance: regardless of the connection with the
optical counterpart, the prompt GW analysis reported a sky-averaged distance
estimate of $39\pm17$ Mpc.

Assuming a distance of 40 Mpc (i.e. to NGC~4993; \cite{Freedman01}), the
\emph{Fermi}-GBM detection corresponds to an isotropic-equivalent 1 keV--10 MeV
luminosity $L_{\rm iso}=1.6 (\pm0.6)\times10^{47}$\ erg s$^{-1}$ \cite{LVCGBM}.
This is three orders of magnitude below the lowest-energy sGRB in the sample
reported by \cite{Davanzo14}, consistent with the predictions of \cite{Evans16}.
That work compared the observed flux of sGRBs detected by {\it Swift},
\emph{Fermi} and \emph{CGRO} without known redshift with that which would have
been seen from the {\it Swift}\ sGRBs of known redshift had they been in the
range of advanced LIGO/Virgo. They predicted that a binary neutron star merger
detected by advanced LIGO/Virgo would be several orders of magnitude less
luminous than GRBs detected previously. Given this low luminosity, GW 170817
allows us to probe the parameter space not covered by the sGRBs detected
hitherto. While the GBM spectral properties of the object \cite{LVCC21528,LVCGBM} are
fairly typical of sGRBs, it should be borne in mind that this event is so
underluminous compared to the previous sample of GRBs, that models constructed
for those objects may not be simply transferrable to such a low-luminosity event
as this. Indeed (see Section~\ref{sec:cocoon}), the low-luminosity is difficult
to explain as arising from an ultra-relativistic jet at all. The implications of
the sub-luminous gamma-ray emission are explored in further detail in
\cite{Kasliwal17}.

\subsection{UV/Optical Emission}

\subsubsection{SED Construction and Fitting}

Since our UVOT detections extend out to only $\Delta t = 1.0$\,d, we focus on the 
early-time behavior of the optical counterpart in this work.  We constructed
the spectral energy distribution (SED) of EM\,170817 using data from our 
work and the GCN Circulars.  We limit the  dataset considered here
to the $izy$ photometry provided by the Pan-STARRS project \cite{LVCC21617,Smartt17}. 
At the time of the preparation of this manuscript, this was one of the few
datasets with host-subtracted photometry (like the UVOT), and the transient
field was previously calibrated in these filters.  We linearly
interpolated the Pan-STARRS 1 observations to match the two epochs of UVOT 
photometry.

After correcting for Galactic extinction, we fit the resulting SED to a
blackbody function.  Adding a free parameter for host extinction did not
improve the fit quality.  At $\Delta t = 0.6$\,d, we find $T_{\mathrm{BB}}
= 7300 \pm 200$\,K, corresponding to $R_{\mathrm{BB}} = 6.0 \times
10^{14}$\,cm and $L_{\mathrm{BB}} = 7 \times 10^{41}$\,erg\,s$^{-1}$.
The blackbody model yields a fit statistic of $\chi^{2}_{\mathrm{r}} = 
2.8$ (5 degrees of freedom).
At $\Delta t = 1.0$\,d, we derive $T_{\mathrm{BB}}
= 6400 \pm 200$\,K, corresponding to $R_{\mathrm{BB}} = 6.5 \times
10^{14}$\,cm and $L_{\mathrm{BB}} = 5 \times 10^{41}$\,erg\,s$^{-1}$.
At this epoch, we derive a goodness of fit of $\chi^{2}_{\mathrm{r}} =
1.6$ (4 degrees of freedom).
The resulting SED fits are plotted in Figure 2B--C.


For comparison, we also fit the SED at both of these epochs to a power-law
model (as would be expected from synchrotron afterglow emission) of the
form $f_{\lambda} \propto \lambda^{-\alpha}$.  At $\Delta t = 0.6$\,d,
we find $\alpha = 1.0 \pm 0.4$.  The fit quality is extremely
poor: $\chi^{2}_{\mathrm{r}} = 102.8$ (5 degrees of freedom).
Repeating this analysis at $\Delta t = 1.0$\,d, we find $\alpha = 1.0
\pm 0.7$ and $\chi^{2}_{\mathrm{r}} = 93.5$ (4 degrees of freedom).
Clearly the power-law model provides a much worse fit to the early
SED.

\subsubsection{Model Comparisons to the Early UV/Optical Light Curve}

Our kilonova  calculations to model the early ($\Delta t \leq
2$\,d) UV/optical emission from EM\,170817 are based on the approach of 
Wollaeger et al. \cite{wkf+17}.  We employ the multigroup, multidimensional 
radiative Monte Carlo code 
{\sc SuperNu}\cite{wollaeger13,wollaeger14a,vanRossum16} with the set of opacities produced 
by the Los Alamos suite of atomic physics codes
\cite{fontes15a,fontes15b,fontes17a}.

We studied a range of 1-dimensional and 2-dimensional 
models, systematically exploring different uncertainties affecting the 
kilonova light curves: ejecta mass, velocity, composition, and morphology, as 
well as the model for the energy deposition in post-nucleosynthetic 
radioactive decays. Our 2-dimensional models are based on the dynamical ejecta 
morphologies from \cite{rosswog14a}, computed by
following long-term evolutions of the ejecta from neutron star mergers, which
in turn were simulated in \cite{rosswog13a}.  The $r$-process 
nucleosynthesis and radioactive heating are computed using the nuclear network 
code {\sc WinNet} \cite{winteler12a,korobkin12,thielemann11}. 
Reaction rates for nucleosynthesis are taken from the compilation of
\cite{rauscher00a} for the finite range droplet model (FRDM; 
\cite{moeller95a}), including density-dependent weak reaction rates 
\cite{arcones11a} and fission \cite{panov10a,panov05}. Coordinate- and 
time-dependent thermalization of nuclear energy is computed using empirical 
fits developed in \cite{barnes16a} and \cite{rosswog17a}.

The abundances of trans-lead elements depend sensitively on the
nuclear mass model and, since they dominate the production of
$\alpha$-particles, the production of these elements can dramatically alter
the heating at late times \cite{rosswog17a,barnes16a}.
On the other hand, the nuclear heating rate on the timescale up to one to two
days is dominated by $\beta$-decay, which shows consistency across different
nuclear mass models \cite{barnes16a}.  Because the modeling here is focused 
on this early-time data, we opt to utilize the FRDM nuclear mass model and do 
not study the uncertainties  of the nuclear heating here.
 
Instead, we focus on the properties of the ejecta, which consist of
two components: a dynamical ejecta component and a wind ejecta component.
The eight most sophisticated models in \cite{wkf+17} have labels of the form 
$\gamma A_1, \gamma A_2, \gamma B_1, \gamma B_2 \ldots \gamma D_1, \gamma D_2$,
where the letter A--D represents the tidal morphology, are taken from \cite{rosswog14a}, see also 
\cite{grossman14}, their figure 1 for an illustration; the subscripted number
represents the wind type (below).

The dynamical ejecta are in an axisymmetric, very neutron-rich outflow
($Y_e\approx0.04$) composed almost entirely of the main $r$-process 
elements with significant mass fraction of lanthanides and actinides 
($\approx 14\%$), taken from \cite{rosswog14a}.  The wind ejecta are in a 
spherically-symmetric, moderately neutron-rich outflow with two different 
degrees of neutron richness: i) (subscript 1 in the model descriptor) $Y_e=0.37$ with a composition dominated 
by iron peak and slightly-above iron peak isotopes, and ii) (subscript 2 in the model descriptor) $Y_e=0.27$ with a 
composition dominated by elements at or just above the first $r$-process 
peak.  Astrophysical and thermodynamical conditions to generate the wind 
compositions were sampled from typical outcomes of the high- and 
low-latitude wind trajectories, simulated in \cite{perego14a} and
\cite{martin15}.

The $\gamma$-models in \cite{wkf+17} take into account the distinct composition 
of the two outflow components, which in turn determine the opacities, 
ionization conditions, radioactive nuclear heating and thermalization. The 
``$\gamma$'' notation indicates that the gray $\gamma$-ray transport is also 
included to accurately account for the $\gamma$-ray energy escape or 
deposition.

\begin{figure}
 \begin{tabular}{cc}
 \includegraphics[width=0.48\textwidth]{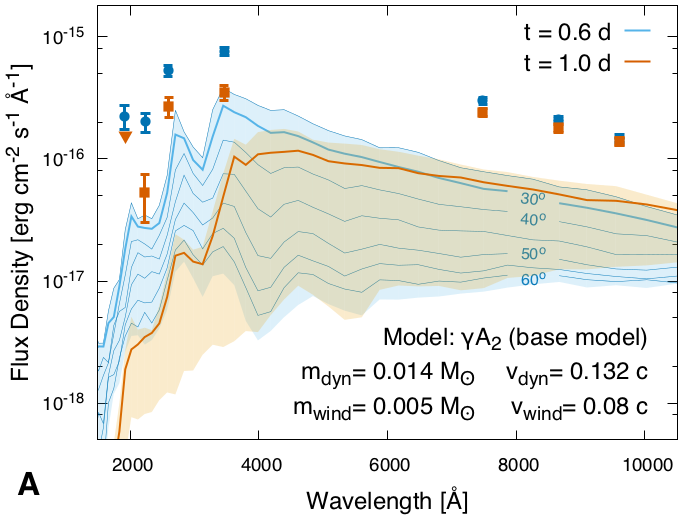} &
 \includegraphics[width=0.48\textwidth]{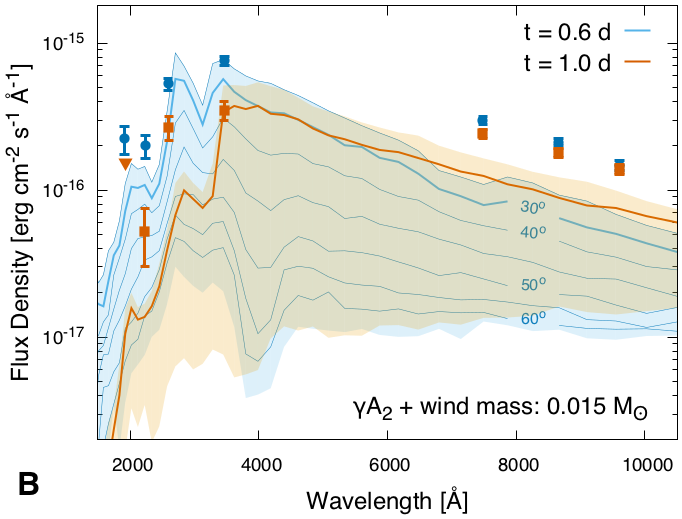} \\
 \includegraphics[width=0.48\textwidth]{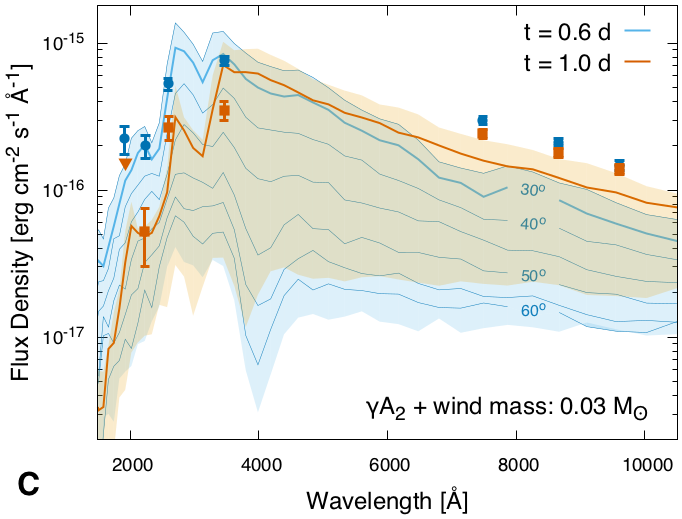} &
 \includegraphics[width=0.48\textwidth]{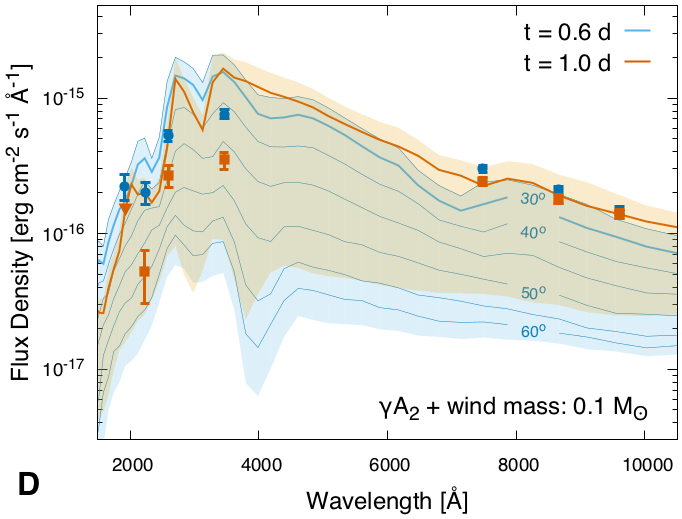} \\
 \end{tabular}
 \caption{{\bf Model SEDs for kilonovae compared to our data.}
 The SEDs were created  for four models based on the $\gamma A_2$ (i.e.\ with the wind-2
 composition), for different orientations and for two different epochs.
 The shaded regions display the range of model SEDs for different viewing angles
 over the range considered here.
 Thick solid lines show the flux for viewing angle
 $\theta\approx30^\circ$, and thin lines represent adjacent angular bins with
 increasing viewing angle going up to $\approx60^\circ$.
 Data points are shown as circles with error bars, or triangles (for upper limits).
 The models span a range of wind masses: a baseline model $\gamma A_2$
 (panel A) has relatively low wind mass of $0.005\ M_\odot$. In the following
 panels (B--D): $m_{\rm wind}=0.015\ M_\odot$,
 $0.03\ M_\odot$ and $0.1\ M_\odot$. To fit the observed data points,
 the wind mass needs to be at least $0.03\ M_\odot$.}
 \label{fig:windmass}
\end{figure}

To fit the observed data, we allowed the mass and velocity to vary for both of
our two components. We also considered two different compositions for
the wind ejecta based on electron fractions of 0.27 and 0.37.  The SEDs at times
of $0.6$\,d and $1.0$\,d after the launch of the explosion are shown in 
Figure~\ref{fig:windmass}.  This figure plots the flux as a 
function of wavelength for a variety of viewing angles. We focused on 
the dynamical ejecta 
configuration A, because only in this morphology the polar region is free 
from the ``lanthanide curtain'' created by the dynamical ejecta, and the wind is 
completely unobscured when viewed on-axis.

Figure~\ref{fig:windmass} demonstrates that in order to fit the observed
emission, we must raise the wind ejecta mass to at least $0.03\ M_\odot$ (see
panels C--D). With a wind ejecta mass of $0.1\
M_\odot$, a fit to the data can be made with a viewing angle of 40$^\circ$.
Although this is at the high end of the
predicted wind ejecta masses, it is not impossible to produce such high-mass
accretion disk outflow with e.g. an asymmetric merger 
\cite{Just15,Fernandez15b,Siegel17}. It is also possible
that we are observing more irregular wind morphologies which will in any case
produce brighter transients (by a factor of a few) than our spherical models.

\begin{figure}
 \begin{tabular}{cc}
 \includegraphics[width=0.48\textwidth]{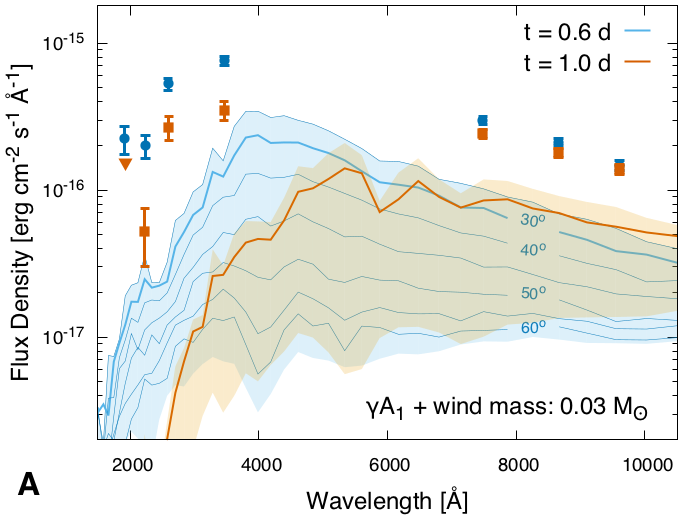} &
 \includegraphics[width=0.48\textwidth]{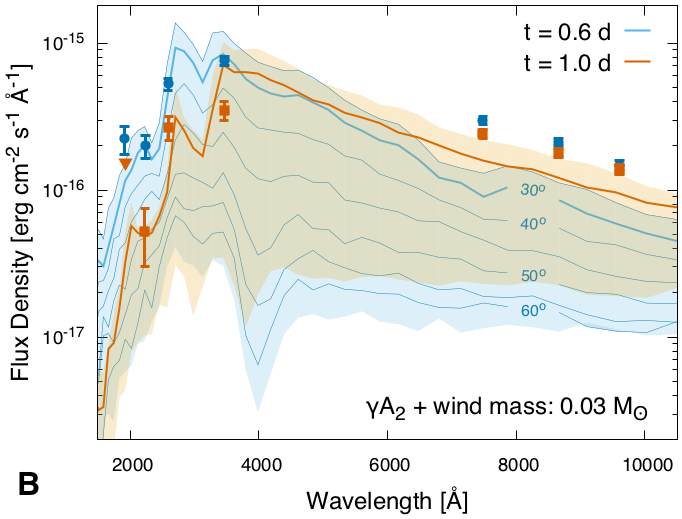} \\
 \end{tabular}
 \caption{{\bf The effect of wind electron fraction on the SED.}
 These SEDs have the same velocities and masses of the
 ejecta but different composition: ``wind 1'' (A) with abundant iron-group
 and the $d$-shell elements vs. ``wind 2'' (B) with the first peak elements,
 largely representing the $s$- and $p$-shell elements and relatively fewer
 $d$-shell elements. Notation for the plots is the  same as in the previous
 figure. The iron-group dominated composition not only exhibits lower
 brightness but also shows much more reddening in the spectrum between the two
 epochs. Datapoints are as in Figure~\ref{fig:windmass}.}
 \label{fig:composition}
\end{figure}

\begin{figure}
 \begin{tabular}{cc}
 \includegraphics[width=0.48\textwidth]{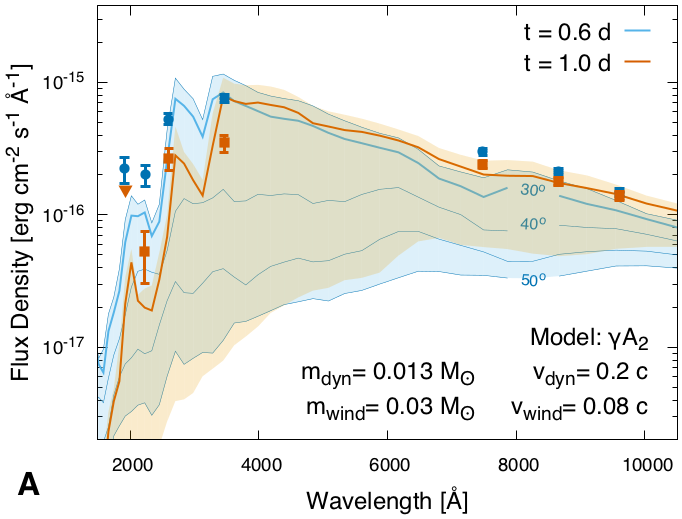} &
 \includegraphics[width=0.48\textwidth]{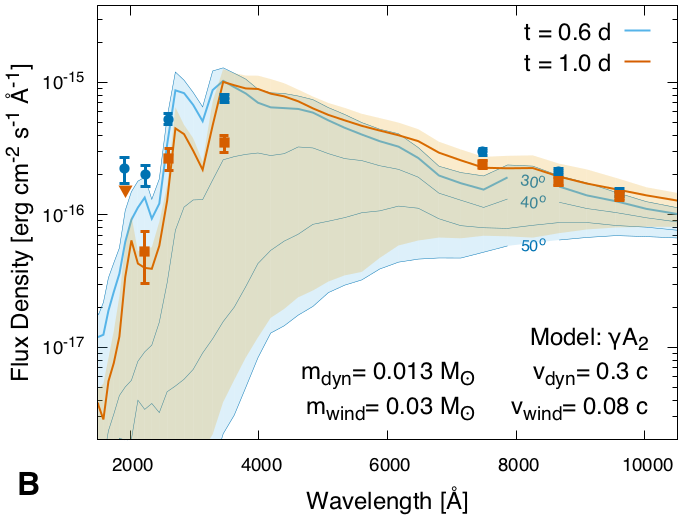} \\
 \end{tabular}
 \caption{{\bf The effect of dynamical ejecta velocity on the SED.}
 These SEDs have the same velocities and masses of the
 wind ejecta but different velocities of the dynamical ejecta: ${v_{\rm
 dyn}=0.2\ c}$ (A) vs. $0.3\ c$ (B). Notation for the plots is the  same
 as in the previous Figure. Dynamical ejecta need to be expanding with at
 least $0.2-0.3\ c$ to be responsible for the red excess around $7000$-$9000$
 \AA. Datapoints are as in Figure~\ref{fig:windmass}.}
 \label{fig:finalfit}
\end{figure}

Figure~\ref{fig:composition} compares the SEDs for wind ejecta mass
${m_{\rm wind}=0.03\ M_\odot}$ between the two different wind compositions that we
explore. The high electron-fraction wind
ejecta model is unable to match the data without increasing the ejecta mass
further, favoring of the neutron-rich wind ejecta, which indeed is close
to the values predicted for the polar regions \cite{Siegel17,perego14,martin15}.
The spectrum for the model with high electron-fraction exhibits much faster reddening
between the epochs than its medium electron-fraction counterpart.

We emphasize that the ``effective'' opacity
of our wind model 2 ($Y_{e} = 0.27$; first $r$-process peak elements) is actually
somewhat lower than that of wind model 1 ($Y_{e} = 0.37$; Fe-peak elements with
effective $\kappa \approx 1$\,cm$^{2}$\,g$^{-1}$).  This is a result of the 
less complex atomic configurations \cite{wkf+17}.  As a result, the models with
the Fe-peak composition for the disk wind become redder much more quickly and
are therefore difficult to match with the bright UV emission seen at early 
times.

The neutron rich dynamical ejecta does not directly contribute significantly
to the ultraviolet and optical emission in the first two days. In fact, the
spectra and light curves in optical and UV are completely insensitive to the
dynamical ejecta mass, and only weakly sensitive to its velocity in the red
part of the optical spectrum. Figure~\ref{fig:finalfit} represents our best
models in which we have increased the dynamical ejecta velocity to produce the
early red excess in the three points around 7000--8000\,\AA\ (note that velocities
here refer to the average speed of the ejecta; the fastest-moving material has
a velocity of twice this value).
Faster dynamical ejecta ignite the infrared transient earlier, and as a
result causes moderate enhancement in the red part of the spectra. 
We can see that velocity $v_{\rm dyn}=0.2-0.3\ c$ is compatible
with observations. Faster wind, however, causes the blue transient to expire 
sooner and shift the spectrum redwards earlier, which apparently excludes wind 
velocities as high as $v_{\rm dyn}=0.16\ c$.

\begin{figure}
 \begin{tabular}{ccc}
 \includegraphics[width=0.324\textwidth]{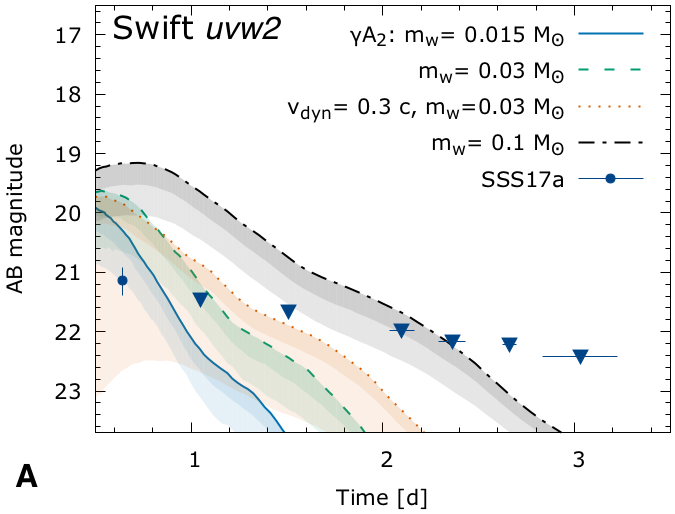} &
 \includegraphics[width=0.324\textwidth]{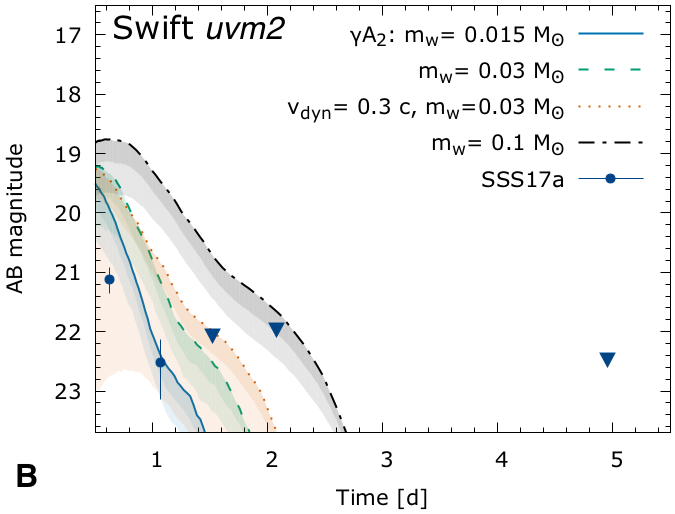} &
 \includegraphics[width=0.324\textwidth]{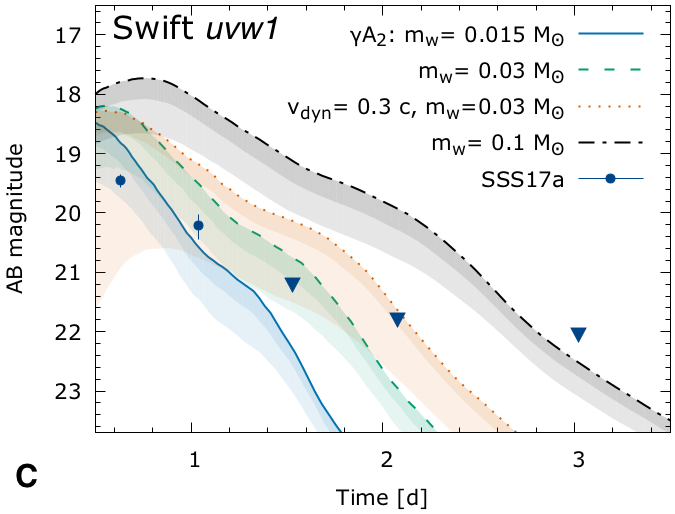} \\
 \includegraphics[width=0.324\textwidth]{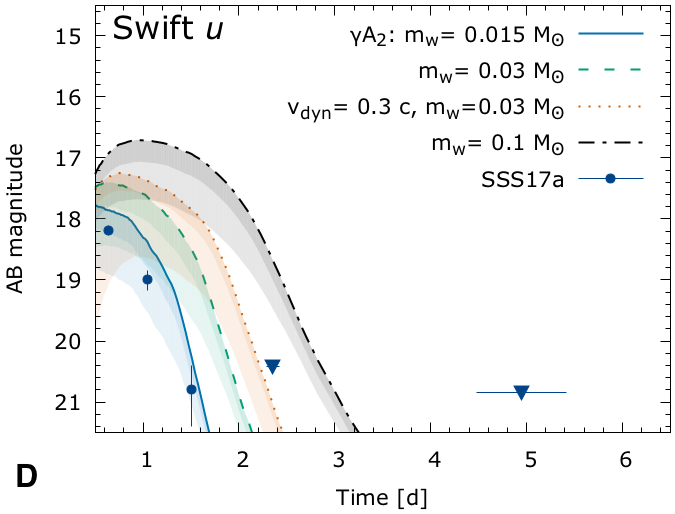} &
 \includegraphics[width=0.324\textwidth]{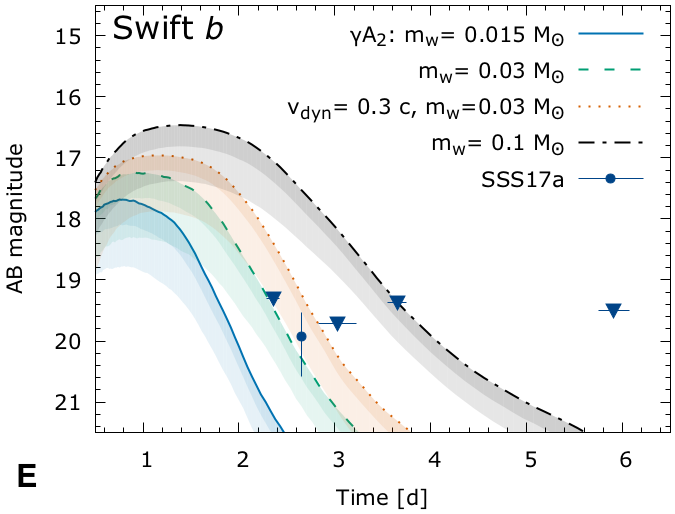} &
 \includegraphics[width=0.324\textwidth]{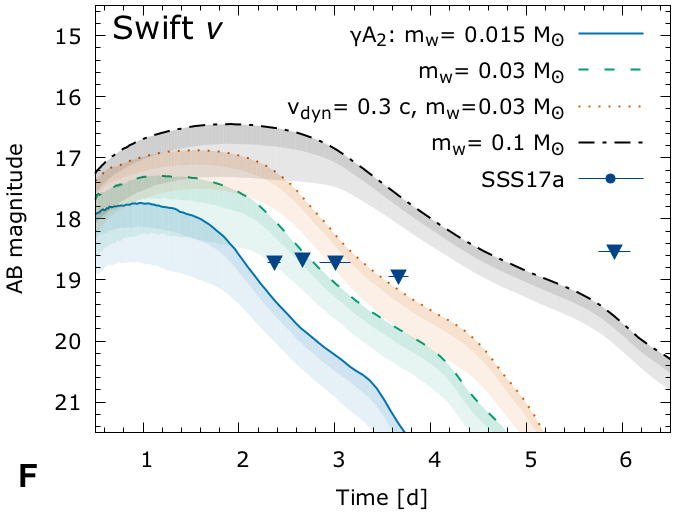} \\
 \end{tabular}
 \caption{{\bf Model light curves for the variants of model configuration $\gamma A_2$}. The solid lines
 indicate the predictions for an on-axis viewing geometry, and 
 the shades beaneath are for  off-axis geometries, with limiting
 angles $30^\circ$ (darker shade) and $40^\circ$ (lighter shade). Datapoints are as in Figure~\ref{fig:windmass}}
\label{fig:uvbands}
\end{figure}

Although the direct contribution of the dynamical ejecta is small, the lanthanides
and actinides in its composition can obscure any emission and, for our
morphology of the tidal ejecta, the emission in the UV and optical is too
obscured for viewing angles more than 40$^{\circ}$ off axis. If the viewing
angle is greater than this value, the neutron-rich dynamical ejecta must be more
constrained to the orbital plane. In another scenario an asymmetric merger may
end up with `kidney-shaped' dynamical ejecta, which only cover a finite
sector in azimutal angle (as in \cite{grossman14}, their figure 1). If the `kidney' is
facing away from us, it will not obscure the brighter and bluer view of the
wind. In either case, it seems that the favored scenario is the one in which
the dynamical ejecta do not strongly obscure the wind.

We conclude with the following optimal range of parameters:
\begin{itemize}
 \item wind composition: first peak elements;
 \item wind mass: $m_{\rm wind} = 0.03 - 0.1\ M_\odot$;
 \item wind velocity: $v_{\rm wind} = 0.08\ c$;
 \item wind kinetic energy: $E_{\rm wind} = 2\times10^{50}$ erg
 \item dynamical ejecta mass: poorly constrained, compatible with the range
       from $m_{\rm dyn} = 0.002 - 0.03\ M_\odot$;
 \item dynamical ejecta velocity: $v_{\rm dyn}=0.2-0.3\ c$;
 \item dynamical ejecta kinetic energy: $E_{\rm dyn}\sim 6\times10^{50}$ erg;
 
 \item viewing angle: $<\sim40^\circ$, degenerate with the wind outflow mass: higher polar angle
       implies higher mass, or non-axisymmetric configuration without dynamical
       ejecta obscuring the wind.
\end{itemize}

Figure~\ref{fig:uvbands} presents the light curves in the UV bands for the
variations of our baseline model $\gamma A_2$.

In addition to the early data from {\it Swift}-UVOT, the Very Large Telescope (VLT) in Chile observed
EM\,170817 with the U filter (similar to the $u$ filter on UVOT) on the Visible wide field Imager and Multi-Object Spectrograph
(VIMOS) instrument at $\Delta t=4.47$ d. Observations were processed via the {\tt esorex} software
and the photometric zero-point was established using the UVOT data for field stars, transformed to the VLT U-band
using the relationsi n \cite{Poole08}. The (extinction corrected)
magnitude at this time was $23.35 \pm 0.07$ (AB). Similar results were found by the Hubble Space Telescope \cite{Kasliwal17}.
These points are not explained by our model, which predicts no emission at these wavelengths at late times
(Figure~\ref{fig:uvbands}D). However, since our focus is on early-time emission certain
assumptions were made above (such as neglecting uncertainties on nuclear heating).
Thus our which focusses only on the early-time data and reproduced this well,
does not give a complete picture of the emission processes at late times.

\subsection{X-ray Emission}

\subsubsection{Afterglow emission}
\label{sec:xon}

To place our X-ray limits at the location of EM\,170817 in context, we simulated a number of fake sGRB afterglow light
curves at a distance of 40 Mpc. To do this we took the sample of sGRBs with known redshift from \cite{Davanzo14}, for
each of these the best-fitting power-law models from the online XRT GRB
Catalog (http://www.swift.ac.uk/xrt\_live\_cat) \cite{Evans09}. Occasionally we manually added extra power-law segments to more
accurately reproduce the light curves. This was necessary because for the catalogue extra segments are only added when
they are justified at the 4-$\sigma$ level, whereas for our purposes we simply want a model that accurately describes
the data at the time of our observations. We then scaled the normalisation to correct from the observed
redshift to 40 Mpc.

In Figure 4 we plot the resultant median light curve and the 25th and 75th 
percentiles; also shown are the \emph{Swift} and \emph{NuSTAR} upper limits.  While the sample of sGRBs used is clearly biased, 
both in sample definition (it is intended to be complete in terms of BAT 
sensitivity) and our extra selections (objects with no XRT detection cannot be 
included), this nonetheless indicates that the X-ray afterglow of GRB\,170817A 
is significantly fainter than the majority of previously-observed sGRBs. 

This result is not entirely surprising, since the gamma-ray output was orders 
of magnitude less luminous than the current sample, and this is known to 
correlate with the X-ray afterglow luminosity. \cite{Davanzo14} presented 
correlations between the prompt isotropic energy output and luminosity 
($E_{\gamma,\mathrm{iso}}, L_{\gamma,\mathrm{iso}}$) and the X-ray flux at 5 
hours. Using the \emph{Fermi}-GBM values of $E_{\gamma,\mathrm{iso}} = 4.0
\times 10^{46}$\,erg and $L_{\gamma,\mathrm{iso}} = 1.6 \times 
10^{47}$\,erg\,s$^{-1}$ \cite{LVCGBM}, and assuming a power-law decay of $L \propto 
t^{-1.5}$ (based on the median light curve in Figure~4), these correlations 
predict an X-ray luminosity at the time of our first \emph{Swift}-XRT 
observation of $4.5\times10^{40}$ erg s$^{-1}$ (from $E_{\gamma,\mathrm{iso}}$) 
or $2.1\times10^{41}$ erg s$^{-1}$ (from $L_{\gamma,\mathrm{iso}}$). Our upper 
limit at this time was $2.2\times10^{40}$ erg s$^{-1}$. Thus, if the X-ray 
afterglow of GRB\,170817A followed the \cite{Davanzo14} correlations exactly, 
we would have expected to detect it (though there is some uncertainty in the 
correlations).

While X-rays were detected by {\it Chandra} from EM\,170817 at around 9 days after the trigger (no flux was
reported for this proprietary observations) \cite{LVCC21765,Troja17,Haggard17,Margutti17},
extrapolating the detected X-ray flux back to the times of the \emph{Swift} and \emph{NuSTAR} observations using the
$t^{-1.5}$ power-law yields a flux above our upper limits \cite{LVCC21765}. Thus our early non-detections also disfavor
an on-axis afterglow as the source of the late-time X-ray emission detected by {\it Chandra}.


In addition to a comparison with the observed sample of sGRBs, we can 
translate these limits on X-ray emission from EM\,170817 to physical constraints on 
any associated relativistic ejecta using the standard synchrotron afterglow
formulation \cite{Sari98}. First, we consider on-axis geometries (i.e.\ where 
the relativistic jet is pointed towards Earth).

The X-ray flux will depend on the location of the cooling frequency $\nu_c$. At 
the time of our first \emph{Swift}-XRT observation ($\Delta t = 0.6$\,d), the
cooling frequency will be \cite{gs02}:

\begin{equation}
\nu_{c} = 4 \times 10^{18} \left( \frac{n_{0}}{5 \times 10^{-3}\,\mathrm{cm}^{-3}}
\right)^{-1} \left( \frac{E_{\mathrm{AG}}}{2 \times 10^{51}\,\mathrm{erg}}
\right)^{-1/2} \, \mathrm{Hz},
\end{equation}

\noindent where we have normalized the isotropic afterglow energy ($E_{\rm AG}$)
and circumburst density ($n_o$) to the median values for sGRBs derived in
\cite{Fong15}. Thus, the XRT bandpass (0.3--10.0\,keV =
$7.3\times10^{16}$--$2.4\times10^{18}$ Hz) will typically fall near (but
slightly below) the cooling frequency at this time, while $\nu_{c}$ will often
fall within the (broader) \emph{NuSTAR} bandpass. (Here and througout, we assume
a constant-density circumburst medium, with microphysical parameters:
$\epsilon_{e} = 0.1$ and $\epsilon_{B} = 0.01$, and an electron spectral index
of $p = 2.5$.)

For $\nu_{X} < \nu_{c}$, our initial \emph{NuSTAR} upper limit implies 
\cite{gs02}:

\begin{equation}
\left( \frac{n_{0}}{5 \times 10^{-3}\,\mathrm{cm}^{-3}} \right)^{1/2}
\left( \frac{E_{\mathrm{AG}}}{2 \times 10^{51} \, \mathrm{erg}} \right)^{11/8}
< 2 \times 10^{-4}.
\end{equation}

\noindent As a result, for any reasonable circumburst density ($n_{0} > \sim 
10^{-5}$\,cm$^{-3}$; \cite{Fong15}), we place a limit on the amount of energy 
coupled to relativistic ejecta along our line of sight of $E_{\mathrm{AG}} 
<\sim 10^{50}$\,erg.  Adopting $\nu_{X} > \nu_{c}$ yields 
similar constraints (though the result is independent of $n_{0}$).  
Utilizing the first \emph{Swift}-XRT limit results in a comparable limit on
the energy coupled to relativistic ejecta (for on-axis geometries).

To verify these results, we ran a series of simulations using the afterglow 
light curve code {\sc boxfit} \cite{vanEerten12}. Over the observed range of 
circumburst densities and afterglow energies for sGRBs\cite{Fong15}, we 
calculated the predicted X-ray flux at the time of our initial \emph{NuSTAR}
observation ($\Delta t = 0.7$\,d). The results are shown in the bottom
panel of Figure~4.  Similar to our analytic results above, we can rule out 
the presence of a collimated, relativistic outflow initially beamed towards 
Earth with $E_{\mathrm{AG}} >\sim 10^{50}$\,erg.

As discussed by \cite{Kasliwal17}, the prompt low $\gamma$-ray luminosity of
GRB\,170817A is not consistent with emission from a standard ultra-relativistic
jet as seen in sGRBs to date, regardless of viewing angle. Instead it must arise
from some previously unseen process, such as a heated cocoon (e.g.\
\cite{Lazzati17,Gottlieb17}); and we are likely viewing the event off-axis
(unless there was no jet at all), as indeed our modelling of the UV emission
(above) suggests. In this case, if there is an ultrarelativistic jet which is
oriented away from us, we can expect X-ray emission at late times. As the jet
decelerates and spreads laterally, it illuminates an increasing fraction of the
sky, thus eventually spreading into our line of sight. Such ``orphan'' afterglow
emission is expected to be much fainter than the early afterglow, but due to the low distance
to GW\,170817 may be detectable by {\it Swift}.

We therefore repeated the afterglow light curve simulations described above for
a variety of off-axis viewing geometries. For the median values of short-duration GRB
afterglow energy ($E_{\mathrm{AG}} = 2 \times 10^{51}$\,erg) and circumburst
density ($n_{0} = 5 \times 10^{-3}$\,cm$^{-3}$; \cite{Fong15}), the resulting
light curves are plotted in Figure~5. With the XRT and \emph{NuSTAR}
non-detections to date, we can rule out any viewing angle with
$\theta_{\mathrm{obs}} <\sim 20^{\circ}$. With these afterglow parameters, the
reported \emph{Chandra} flux of $5 \times 10^{-15}$\,erg\,cm$^{-2}$\,s$^{-1}$
\cite{LVCC21787,LVCC21798} can be reproduced with a viewing angle of
$\theta_{\mathrm{obs}} \approx 30^{\circ}$. These constraints are consistent
with the geometry implied by our modelling of the blue kilonova seen by the
UVOT. It is also broadly consistent with the evolution of EM\,170817 at radio wavelengths
\cite{Hallinan17}, showing no emission at early times but then later detections
\cite{LVCC21814,LVCC21815}. 


\subsubsection{Emission from a heated cocoon}
\label{sec:cocoon}

One way of explaining the observed prompt gamma-rays may be via the `cocoon
model' \cite{Lazzati17,Gottlieb17}. In this model, if the GRB jet propagates
through a baryon-contaminated site around the merger, a heated cocoon is formed
from which prompt gamma-rays may be seen. Under the \cite{Lazzati17} model, the  prompt
luminosity of GRB 170817A can be interpreted as arising from a typical (or
rather, {\it Swift}-like) sGRB viewed $\sim$10--20$^\circ$ off-axis. In their
revised model, in which the cocoon is not isotropic \cite{Lazzati17b}, the
prompt gamma-ray luminosity implies an observing angle of $\sim$40--50$^\circ$ compared to the
jet angle. However,  the peak photon energy reported by \cite{LVCC21528}
($124.2\pm52.6$ keV) requires a viewing angle of $\sim10^\circ$ according to the
same model, which suggests that this model, in its current form at least, cannot
explain the observed prompt emission.

Nonetheless, a heated cocoon may in principle give rise to late-time X-ray
emission, as the `trans-relativistic' expanding cocoon (i.e.\ with Lorentz
factor $\Gamma\sim$2--3) is eventually decelerated by the ambient medium which
then shocks and radiates in a manner analagous to the standard on-axis GRB
afterglow \cite{Gottlieb17}. In the main text we demonstrated that our upper
limits and the {\it Chandra} detections are consistent with most of these
models, which predict a peak X-ray flux several weeks to months after the
merger. 


\subsubsection{Non-afterglow emission}
\label{sec:oamodels}

In addition to the emission associated directly or indirectly with the jet, as
above, some authors have predicted quasi-isotropic X-ray emission following
a binary neutron star merger. \cite{Kisaka16}, responding to an apparent X-ray excess in the light
curve of GRB 130603B \cite{Fong14} coincident with a claimed kilonova detection
\cite{Tanvir13,Berger13}, proposed that the kilonova was powered not by
\emph{r}-process nucleosynthesis but by isotropic X-ray emission, perhaps from
fallback accretion via a disk. They predict typical X-ray luminosities of order
$10^{41}$ erg s$^{-1}$ appearing at around 7 days after the merger. We find no
evidence for such emission in our data, with upper limits of
$\sim1.5\times10^{40}$ erg s$^{-1}$ at 7 days after the trigger.

{\it Chandra} detected X-ray emission from EM\,170817 at $\sim$ 9 d after the
trigger \cite{LVCC21765} and in subsequent observations
\cite{LVCC21786,LVCC21787,LVCC21798,Troja17,Haggard17,Margutti17} with a luminosity of $\sim9\times10^{38}$
erg s$^{-1}$ (0.3--8 keV), consistent with our XRT and {\it NuSTAR} upper
limits. This could be indicative of the emission predicted by \cite{Kisaka16},
if the accretion rate is much lower than in their fiducial calculations.
Alternatively, those authors note that if the ejecta initially prevent the
detection of the X-rays (i.e.\ the X-rays are all thermalized by the
optically thick ejecta) then the ejecta will still become optically thin to
X-rays at tens of days, with luminosities $\sim10^{40}-10^{41}$ erg s$^{-1}$,
still well above both the XRT/\emph{NuSTAR} detection thresholds and the
detected {\it Chandra} flux, thus testable with future observations.

It has also been considered that the result of the BNS merger may be a
hyper-massive neutron-star which is able to support itself at least temporarily
against gravitational collapse, e.g.\ by rotation, and which emits X-rays as it
rotates. For example \cite{Rowlinson10,Rowlinson13,Gompertz13,Gompertz15} succesfully modelled
early-time X-ray light curve plateaux as being powered by the spin-down of a
magnetar. However such plateaux do not extend to the times at which our
observations of EM\,170817 began, and in some cases imply the collapse of the
magnetar within a few ks of the merger, well before our X-ray observations of
EM\,170817 began.

\cite{Sun17} proposed three mechanisms for producing X-rays after a binary
neutron star merger: the standard afterglow emission, spin-down energy extracted
from a magnetar, or a `merger-nova' in which the kilonova-emitting ejecta are
heated also by the magnetar and emit X-rays (this latter case was
also proposed by \cite{Siegel16} in whose model this emission is extremely bright,
however \cite{Sun17} attribute this to an error and predict dramatically less
X-ray flux). In this model, the observed X-ray emission depends upon the
viewing angle; with the observer either looking down the GRB jet, missing the
jet but having an unobstructed view of the magnetar, or missing the jet, and
viewing the magnetar through the ejecta. Based on these assumptions,
\cite{Sun17} produced a set of example light curves for typical sGRB parameters.
Almost all of their light curves predict luminosities 4--5 orders of magnitude
above the X-ray detection limits presented here, thus even if the available
energy in the magnetar is substatially reduced compared to typical, we would
have detected this emission.

However, the X-ray flux is greatly reduced in the case that the magnetar
collapses to a black hole before our observations (consistent with the plateau
durations of \cite{Rowlinson10,Rowlinson13,Gompertz13,Gompertz15}), or the
collapse timescale of the magnetar is shorter than its spindown timescale. In
light of these considerations, the lack of X-ray emission above
$\sim1.5\times10^{40}$ erg s$^{-1}$ strongly suggests that if a magnetar was
formed after the BNS merger, it collapsed to a black hole within the first
0.6 d of its formation.

\subsubsection{Limits on nuclear line emission}

\begin{figure}
\begin{center}
\includegraphics[width=12.1cm]{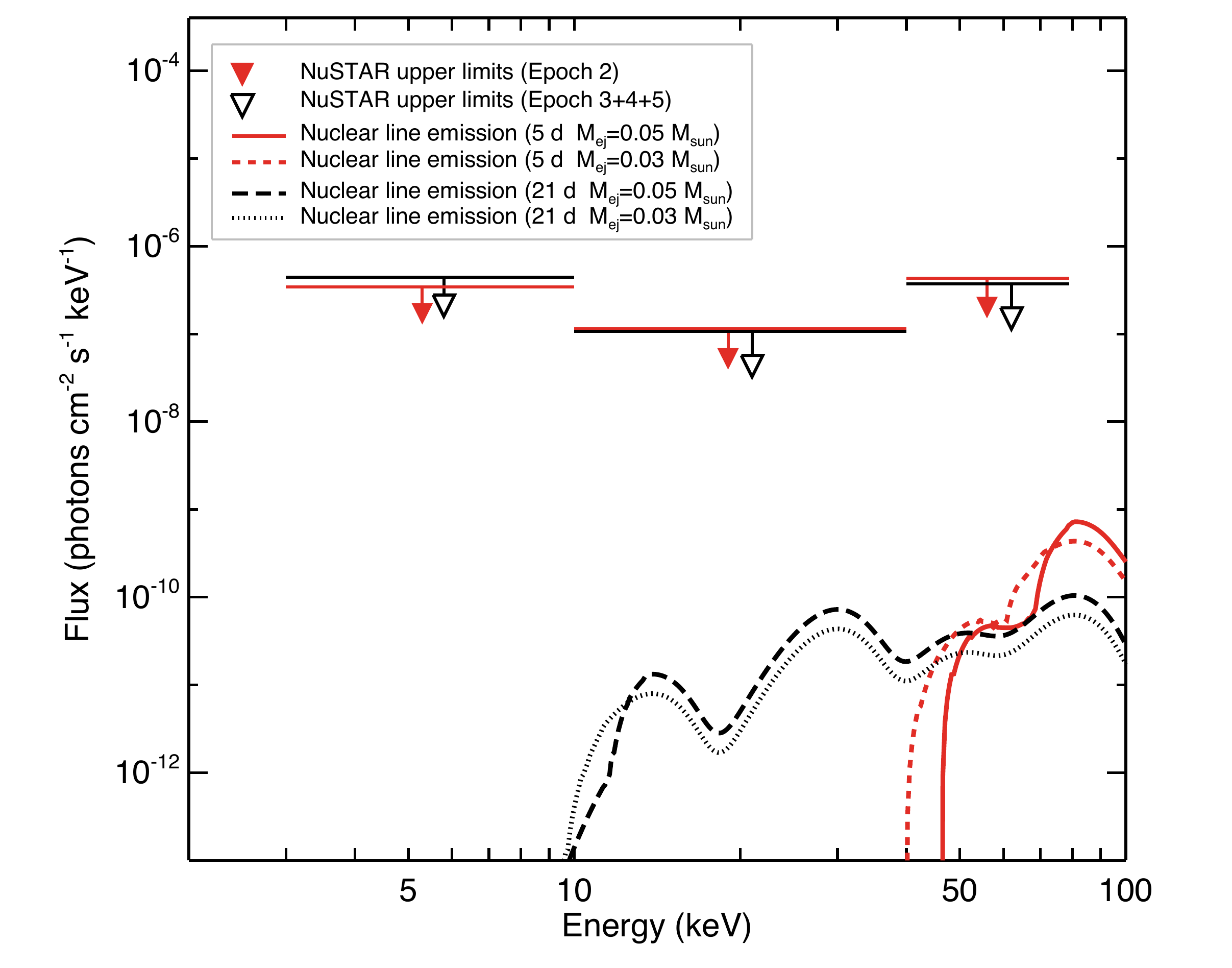}
\end{center}
\caption{
{\bf Predicted X-ray flux from atomic line emission of high atomic mass elements.} The predicted flux
falls well below the sensitivity of our {\it NuSTAR} data.}
\label{fig:lines}
\end{figure}

Radioactive $r$-process nucleii synthesized in the merger ejecta produce gamma-ray lines through $\beta$ decay. The emission
lines have a broad spectrum from 10 keV to 1 MeV \cite{Hotokezaka16,barnes16a}). These lines start to emerge
from the ejecta when the ejecta become optically thin. We model the flux of the nuclear emission lines for the ejecta
mass of 0.03 $M_\odot$ and 0.05 $M_\odot$ with the mean expansion velocity of 0.2$c$ \cite{Hotokezaka16}, taking photoelectric
absorption and the Compton scattering into account. Here we assume a composition of nuclei that matches the solar
$r$-process abundance pattern of stable and long-lived nuclei for $90\leq A \leq 140$, where $A$ is the atomic mass
number. There are three strong lines around $28$, $50$, and $81$ keV, which are produced by $^{129}$Te, $^{132}$Te, and
$^{133}$Xe. However, the line nuclear emission is too weak to be detected by {\it NuSTAR}, as shown in Figure~\ref{fig:lines}.

\subsection{Constraints on X-ray emission from the wide-angle search.}

Our discussion so far has assumed that EM\,170817, GRB 170817A and GW 170817 are
all the same event. While this seems reasonable on probabilistic grounds (e.g.
\cite{LVCC21557,Capstone}) it is not certain. For this reason, even after the
detection of EM\,170817, {\it Swift} continued to search the GW error region for
possible counterparts. In total the tiling observations continued until 4.3 d after the GW
event, and XRT covered 52\%\ of the probability in the three-detector
`prelimary-LALInference' skymap \cite{LVCC21527}, and 92\%\ of the probability determined by convolving
this skymap with GWGC, as described in \cite{Evans16c}. Therefore, if EM\,170817
is not the counterpart to GW 170817 it is probable that {\it Swift}-XRT observed
the true location of the event.

In Figure~4  we show a sample of sGRB X-ray light curves as
they would have appeared at 40 Mpc (the best-fitting luminosity distance in the
3D GW skymap). The blue line on this plot shows the time-range spanned by our
search of the GW error region and the detection limit of the observations. This
shows that a sGRB afterglow typical of those observed to date would have been
detectable with XRT during our follow up. As noted above, some known sGRBs have
extremely fast-fading afterglows \cite{Rowlinson10}; such objects fall below the
25th percentile of GRB flux by the times of our observations so cannot be seen
in Figure~4 and would not have been detected in our search.

To investigate in more detail, we simulated 10,000 sGRBs. Their position and
distance were drawn at random from the 3-D `bayestar-HLV' skymap \cite{LVCC21513} (i.e.\ that on 
which our tiles were based), and their light curves were produced by taking a random
sGRB from the sample described in Section~\ref{sec:xon} and scaling it to the
distance of the simulated event. We then identified for each of these simulated
events whether its location was observed by XRT, and if so, whether the source
would have been above our detection limit at the time of the observation. We
found that in 65\%\ of cases the source would have been detected. We then
repeated the process, but instead of using the actual locations observed by XRT,
which included many observations of EM\,170817 which displaced planned tiling of
the GW-GRB probability region, we used those which would have been observed if
we had not focussed on EM\,170817. In these simulations 68\%\ of the simulated
sources were recovered. Since GRB 170817A was much fainter than typical {\it
Swift}-detected sGRBs, we repeated these simulations again, this time reducing
the predicted XRT flux by a factor of 1,000. In this case we find we would have
detected 11\%\ of simulated events, or 10\%\ if we had not targeted
EM\,170817. The slight decrease when not targeting EM\,170817 -- i.e.\
when covering \emph{more} of the GW error region, arises because the focus on
EM\,170817 results in longer exposures and hence the ability to detect fainter sources
in that field, which in this case outweighs the benefits of searching more
fields at a lower sensitivity.

\clearpage

\begin{table*}
\begin{center}
\caption{{\bf {\it Swift} observations of the error region of GW 170817A.}}
\label{tab:obs}
\begin{tabular}{ccc}
\hline
Pointing direction & Start time$^a$ & Exposure \\
(J2000) & (UTC) & (s) \\
\hline
$11^h 47^m 9.20^s, -39\deg 50^\prime 40.2^{\prime\prime}$  &  Aug 17 at 13:37:07 (0.039) &   117  \\
$11^h 48^m 57.01^s, -39\deg 50^\prime 27.9^{\prime\prime}$  &  Aug 17 at 13:38:13 (0.040) &   80  \\
$11^h 48^m 1.57^s, -40\deg 08^\prime 14.1^{\prime\prime}$  &  Aug 17 at 13:39:19 (0.040) &   85  \\
$11^h 46^m 15.97^s, -40\deg 08^\prime 01.1^{\prime\prime}$  &  Aug 17 at 13:40:24 (0.041) &   90  \\
$11^h 45^m 23.31^s, -39\deg 50^\prime 03.0^{\prime\prime}$  &  Aug 17 at 13:41:30 (0.042) &   125  \\
$11^h 46^m 18.37^s, -39\deg 32^\prime 28.2^{\prime\prime}$  &  Aug 17 at 13:42:35 (0.043) &   87  \\
\hline
\end{tabular}
\end{center}
\begin{flushleft}
$^a$ Observations were 2017, times in parentheses are days since the GW trigger.
\end{flushleft}
\end{table*}

\begin{table*}
\begin{center}
\begin{tabular}{ccc}
\hline
Pointing direction & Start time$^a$ & Exposure \\
(J2000) & (UTC) & (s) \\
\hline
$11^h 48^m 3.97^s, -38\deg 57^\prime 51.7^{\prime\prime}$  &  Aug 17 at 15:12:34 (0.11) &   85  \\
$11^h 49^m 49.45^s, -38\deg 57^\prime 44.9^{\prime\prime}$  &  Aug 17 at 15:13:40 (0.11) &   90  \\
$15^h 02^m 24.24^s, -43\deg 31^\prime 11.6^{\prime\prime}$  &  Aug 17 at 17:16:33 (0.19) &   15  \\
$16^h 14^m 12.96^s, -61\deg 03^\prime 11.2^{\prime\prime}$  &  Aug 17 at 17:17:28 (0.19) &   57  \\
$09^h 16^m 10.08^s, -37\deg 53^\prime 03.5^{\prime\prime}$  &  Aug 17 at 17:50:11 (0.21) &   27  \\
$10^h 30^m 24.24^s, -35\deg 03^\prime 59.0^{\prime\prime}$  &  Aug 17 at 17:51:07 (0.22) &   52  \\
\hline
\end{tabular}
\end{center}
\begin{flushleft}
$^a$ Observations were 2017, times in parentheses are days since the GW trigger.
\end{flushleft}
\end{table*}

\begin{table*}
\begin{center}
\begin{tabular}{ccc}
\hline
Pointing direction & Start time$^a$ & Exposure \\
(J2000) & (UTC) & (s) \\
\hline
$12^h 53^m 50.16^s, -38\deg 30^\prime 37.4^{\prime\prime}$  &  Aug 17 at 18:38:07 (0.25) &   57  \\
$12^h 55^m 14.88^s, -39\deg 26^\prime 58.9^{\prime\prime}$  &  Aug 17 at 18:39:32 (0.25) &   62  \\
$12^h 56^m 52.32^s, -39\deg 26^\prime 58.9^{\prime\prime}$  &  Aug 17 at 18:41:03 (0.25) &   60  \\
$13^h 11^m 52.08^s, -43\deg 31^\prime 11.6^{\prime\prime}$  &  Aug 17 at 18:42:27 (0.25) &   57  \\
$13^h 20^m 13.44^s, -44\deg 08^\prime 46.0^{\prime\prime}$  &  Aug 17 at 18:44:05 (0.25) &   60  \\
$14^h 03^m 18.00^s, -33\deg 30^\prime 03.2^{\prime\prime}$  &  Aug 17 at 18:45:37 (0.25) &   57  \\
\hline
\end{tabular}
\end{center}
\begin{flushleft}
$^a$ Observations were 2017, times in parentheses are days since the GW trigger.
\end{flushleft}
\end{table*}

\begin{table*}
\begin{center}
\begin{tabular}{ccc}
\hline
Pointing direction & Start time$^a$ & Exposure \\
(J2000) & (UTC) & (s) \\
\hline
$12^h 41^m 22.08^s, -13\deg 08^\prime 59.6^{\prime\prime}$  &  Aug 17 at 21:24:03 (0.36) &   115  \\
$12^h 42^m 39.36^s, -13\deg 08^\prime 59.6^{\prime\prime}$  &  Aug 17 at 21:26:28 (0.36) &   112  \\
$12^h 43^m 56.40^s, -13\deg 08^\prime 59.6^{\prime\prime}$  &  Aug 17 at 21:28:53 (0.37) &   117  \\
$12^h 45^m 13.68^s, -13\deg 08^\prime 59.6^{\prime\prime}$  &  Aug 17 at 21:31:18 (0.37) &   120  \\
$12^h 45^m 34.32^s, -13\deg 27^\prime 46.8^{\prime\prime}$  &  Aug 17 at 21:33:43 (0.37) &   117  \\
$12^h 46^m 51.60^s, -13\deg 27^\prime 46.8^{\prime\prime}$  &  Aug 17 at 21:36:08 (0.37) &   120  \\
\hline
\end{tabular}
\end{center}
\begin{flushleft}
$^a$ Observations were 2017, times in parentheses are days since the GW trigger.
\end{flushleft}
\end{table*}

\begin{table*}
\begin{center}
\begin{tabular}{ccc}
\hline
Pointing direction & Start time$^a$ & Exposure \\
(J2000) & (UTC) & (s) \\
\hline
$12^h 46^m 59.52^s, -10\deg 19^\prime 55.6^{\prime\prime}$  &  Aug 18 at 02:13:28 (0.56) &   117  \\
$12^h 45^m 51.36^s, -10\deg 38^\prime 42.7^{\prime\prime}$  &  Aug 18 at 02:15:53 (0.57) &   117  \\
$12^h 46^m 0.96^s, -10\deg 57^\prime 29.9^{\prime\prime}$  &  Aug 18 at 02:18:23 (0.57) &   110  \\
$12^h 44^m 55.44^s, -11\deg 16^\prime 16.7^{\prime\prime}$  &  Aug 18 at 02:20:48 (0.57) &   117  \\
$12^h 43^m 51.12^s, -11\deg 35^\prime 03.8^{\prime\prime}$  &  Aug 18 at 02:23:19 (0.57) &   120  \\
$12^h 45^m 21.60^s, -11\deg 53^\prime 51.0^{\prime\prime}$  &  Aug 18 at 02:25:49 (0.57) &   122  \\
\hline
\end{tabular}
\end{center}
\begin{flushleft}
$^a$ Observations were 2017, times in parentheses are days since the GW trigger.
\end{flushleft}
\end{table*}

\begin{table*}
\begin{center}
\begin{tabular}{ccc}
\hline
Pointing direction & Start time$^a$ & Exposure \\
(J2000) & (UTC) & (s) \\
\hline
$12^h 42^m 52.56^s, -14\deg 24^\prime 08.3^{\prime\prime}$  &  Aug 18 at 06:57:57 (0.76) &   125  \\
$12^h 43^m 45.12^s, -14\deg 05^\prime 21.1^{\prime\prime}$  &  Aug 18 at 07:00:27 (0.76) &   125  \\
$12^h 45^m 2.64^s, -14\deg 05^\prime 21.1^{\prime\prime}$  &  Aug 18 at 07:02:58 (0.77) &   120  \\
$12^h 45^m 27.60^s, -14\deg 24^\prime 08.3^{\prime\prime}$  &  Aug 18 at 07:05:23 (0.77) &   117  \\
$12^h 45^m 54.00^s, -14\deg 42^\prime 55.4^{\prime\prime}$  &  Aug 18 at 07:07:48 (0.77) &   112  \\
$12^h 46^m 45.12^s, -14\deg 24^\prime 08.3^{\prime\prime}$  &  Aug 18 at 07:10:13 (0.77) &   120  \\
\hline
\end{tabular}
\end{center}
\begin{flushleft}
$^a$ Observations were 2017, times in parentheses are days since the GW trigger.
\end{flushleft}
\end{table*}

\begin{table*}
\begin{center}
\begin{tabular}{ccc}
\hline
Pointing direction & Start time$^a$ & Exposure \\
(J2000) & (UTC) & (s) \\
\hline
$12^h 48^m 8.88^s, -10\deg 01^\prime 08.4^{\prime\prime}$  &  Aug 18 at 10:09:03 (0.89) &   125  \\
$12^h 48^m 15.84^s, -10\deg 19^\prime 55.6^{\prime\prime}$  &  Aug 18 at 10:11:33 (0.90) &   120  \\
$12^h 49^m 40.80^s, -10\deg 38^\prime 42.7^{\prime\prime}$  &  Aug 18 at 10:13:57 (0.90) &   120  \\
$12^h 49^m 50.64^s, -10\deg 57^\prime 29.9^{\prime\prime}$  &  Aug 18 at 10:16:29 (0.90) &   115  \\
$12^h 51^m 7.20^s, -10\deg 57^\prime 29.9^{\prime\prime}$  &  Aug 18 at 10:18:54 (0.90) &   112  \\
$12^h 50^m 1.92^s, -11\deg 16^\prime 16.7^{\prime\prime}$  &  Aug 18 at 10:21:19 (0.90) &   115  \\
\hline
\end{tabular}
\end{center}
\begin{flushleft}
$^a$ Observations were 2017, times in parentheses are days since the GW trigger.
\end{flushleft}
\end{table*}

\begin{table*}
\begin{center}
\begin{tabular}{ccc}
\hline
Pointing direction & Start time$^a$ & Exposure \\
(J2000) & (UTC) & (s) \\
\hline
$12^h 53^m 20.64^s, -15\deg 20^\prime 29.8^{\prime\prime}$  &  Aug 18 at 13:20:09 (1.0) &   120  \\
$12^h 53^m 52.08^s, -15\deg 39^\prime 16.6^{\prime\prime}$  &  Aug 18 at 13:22:33 (1.0) &   115  \\
$12^h 54^m 24.96^s, -15\deg 58^\prime 03.7^{\prime\prime}$  &  Aug 18 at 13:24:59 (1.0) &   117  \\
$12^h 54^m 59.28^s, -16\deg 16^\prime 50.9^{\prime\prime}$  &  Aug 18 at 13:27:23 (1.0) &   112  \\
$13^h 38^m 11.04^s, -36\deg 00^\prime 20.5^{\prime\prime}$  &  Aug 18 at 13:52:35 (1.0) &   70  \\
$14^h 00^m 9.12^s, -48\deg 12^\prime 58.7^{\prime\prime}$  &  Aug 18 at 13:54:40 (1.1) &   117  \\
\hline
\end{tabular}
\end{center}
\begin{flushleft}
$^a$ Observations were 2017, times in parentheses are days since the GW trigger.
\end{flushleft}
\end{table*}

\begin{table*}
\begin{center}
\begin{tabular}{ccc}
\hline
Pointing direction & Start time$^a$ & Exposure \\
(J2000) & (UTC) & (s) \\
\hline
$13^h 07^m 49.44^s, -22\deg 32^\prime 33.7^{\prime\prime}$  &  Aug 18 at 16:59:03 (1.2) &   117  \\
$13^h 23^m 19.92^s, -32\deg 14^\prime 55.0^{\prime\prime}$  &  Aug 18 at 17:01:34 (1.2) &   117  \\
$13^h 56^m 49.44^s, -48\deg 31^\prime 45.8^{\prime\prime}$  &  Aug 18 at 17:04:17 (1.2) &   117  \\
$14^h 15^m 36.96^s, -53\deg 51^\prime 07.2^{\prime\prime}$  &  Aug 18 at 17:07:10 (1.2) &   117  \\
$12^h 41^m 42.00^s, -09\deg 42^\prime 21.2^{\prime\prime}$  &  Aug 18 at 18:00:01 (1.2) &   100  \\
$13^h 11^m 7.20^s, -23\deg 47^\prime 42.4^{\prime\prime}$  &  Aug 18 at 18:02:06 (1.2) &   112  \\
\hline
\end{tabular}
\end{center}
\begin{flushleft}
$^a$ Observations were 2017, times in parentheses are days since the GW trigger.
\end{flushleft}
\end{table*}

\begin{table*}
\begin{center}
\begin{tabular}{ccc}
\hline
Pointing direction & Start time$^a$ & Exposure \\
(J2000) & (UTC) & (s) \\
\hline
$12^h 56^m 18.48^s, -19\deg 43^\prime 29.6^{\prime\prime}$  &  Aug 18 at 21:21:43 (1.4) &   117  \\
$12^h 54^m 58.80^s, -19\deg 43^\prime 29.6^{\prime\prime}$  &  Aug 18 at 21:24:14 (1.4) &   112  \\
$12^h 54^m 30.72^s, -20\deg 02^\prime 16.8^{\prime\prime}$  &  Aug 18 at 21:26:39 (1.4) &   115  \\
$12^h 54^m 4.08^s, -20\deg 21^\prime 03.6^{\prime\prime}$  &  Aug 18 at 21:29:04 (1.4) &   120  \\
$12^h 53^m 10.80^s, -20\deg 02^\prime 16.8^{\prime\prime}$  &  Aug 18 at 21:31:29 (1.4) &   122  \\
$12^h 53^m 38.88^s, -19\deg 43^\prime 29.6^{\prime\prime}$  &  Aug 18 at 21:33:58 (1.4) &   112  \\
\hline
\end{tabular}
\end{center}
\begin{flushleft}
$^a$ Observations were 2017, times in parentheses are days since the GW trigger.
\end{flushleft}
\end{table*}

\begin{table*}
\begin{center}
\begin{tabular}{ccc}
\hline
Pointing direction & Start time$^a$ & Exposure \\
(J2000) & (UTC) & (s) \\
\hline
$13^h 16^m 16.56^s, -26\deg 17^\prime 59.3^{\prime\prime}$  &  Aug 19 at 03:59:38 (1.6) &   125  \\
$13^h 14^m 50.40^s, -25\deg 59^\prime 12.1^{\prime\prime}$  &  Aug 19 at 04:02:09 (1.6) &   122  \\
$13^h 13^m 26.88^s, -25\deg 59^\prime 12.1^{\prime\prime}$  &  Aug 19 at 04:04:39 (1.6) &   120  \\
$13^h 12^m 3.12^s, -25\deg 59^\prime 12.1^{\prime\prime}$  &  Aug 19 at 04:07:04 (1.6) &   115  \\
$13^h 13^m 26.16^s, -25\deg 40^\prime 25.0^{\prime\prime}$  &  Aug 19 at 04:09:29 (1.6) &   127  \\
$13^h 28^m 42.00^s, -31\deg 56^\prime 07.8^{\prime\prime}$  &  Aug 19 at 04:12:00 (1.6) &   117  \\
\hline
\end{tabular}
\end{center}
\begin{flushleft}
$^a$ Observations were 2017, times in parentheses are days since the GW trigger.
\end{flushleft}
\end{table*}

\begin{table*}
\begin{center}
\begin{tabular}{ccc}
\hline
Pointing direction & Start time$^a$ & Exposure \\
(J2000) & (UTC) & (s) \\
\hline
$13^h 00^m 37.68^s, -15\deg 01^\prime 42.6^{\prime\prime}$  &  Aug 19 at 07:10:52 (1.8) &   115  \\
$12^h 59^m 19.92^s, -15\deg 01^\prime 42.6^{\prime\prime}$  &  Aug 19 at 07:13:17 (1.8) &   112  \\
$12^h 58^m 50.88^s, -14\deg 42^\prime 55.4^{\prime\prime}$  &  Aug 19 at 07:15:42 (1.8) &   117  \\
$13^h 00^m 8.64^s, -14\deg 42^\prime 55.4^{\prime\prime}$  &  Aug 19 at 07:18:07 (1.8) &   120  \\
$13^h 08^m 7.44^s, -24\deg 25^\prime 16.7^{\prime\prime}$  &  Aug 19 at 07:20:31 (1.8) &   117  \\
$13^h 36^m 12.72^s, -34\deg 07^\prime 37.6^{\prime\prime}$  &  Aug 19 at 07:23:15 (1.8) &   117  \\
\hline
\end{tabular}
\end{center}
\begin{flushleft}
$^a$ Observations were 2017, times in parentheses are days since the GW trigger.
\end{flushleft}
\end{table*}

\begin{table*}
\begin{center}
\begin{tabular}{ccc}
\hline
Pointing direction & Start time$^a$ & Exposure \\
(J2000) & (UTC) & (s) \\
\hline
$12^h 56^m 44.40^s, -20\deg 21^\prime 03.6^{\prime\prime}$  &  Aug 19 at 10:24:35 (1.9) &   125  \\
$12^h 56^m 19.20^s, -20\deg 39^\prime 50.8^{\prime\prime}$  &  Aug 19 at 10:27:05 (1.9) &   115  \\
$12^h 55^m 55.44^s, -20\deg 58^\prime 37.9^{\prime\prime}$  &  Aug 19 at 10:29:30 (1.9) &   115  \\
$13^h 04^m 1.20^s, -22\deg 13^\prime 46.6^{\prime\prime}$  &  Aug 19 at 10:31:55 (1.9) &   122  \\
$13^h 31^m 45.60^s, -33\deg 30^\prime 03.2^{\prime\prime}$  &  Aug 19 at 10:34:28 (1.9) &   115  \\
$14^h 08^m 10.32^s, -48\deg 31^\prime 45.8^{\prime\prime}$  &  Aug 19 at 10:37:15 (1.9) &   117  \\
\hline
\end{tabular}
\end{center}
\begin{flushleft}
$^a$ Observations were 2017, times in parentheses are days since the GW trigger.
\end{flushleft}
\end{table*}

\begin{table*}
\begin{center}
\begin{tabular}{ccc}
\hline
Pointing direction & Start time$^a$ & Exposure \\
(J2000) & (UTC) & (s) \\
\hline
$14^h 18^m 30.72^s, -55\deg 43^\prime 50.2^{\prime\prime}$  &  Aug 19 at 15:27:30 (2.1) &   117  \\
$13^h 02^m 24.00^s, -22\deg 32^\prime 33.7^{\prime\prime}$  &  Aug 19 at 16:37:27 (2.2) &   112  \\
$13^h 03^m 30.48^s, -22\deg 51^\prime 20.9^{\prime\prime}$  &  Aug 19 at 16:39:53 (2.2) &   117  \\
$13^h 02^m 9.12^s, -22\deg 51^\prime 20.9^{\prime\prime}$  &  Aug 19 at 16:42:24 (2.2) &   117  \\
$13^h 01^m 2.64^s, -22\deg 32^\prime 33.7^{\prime\prime}$  &  Aug 19 at 16:44:49 (2.2) &   122  \\
$12^h 59^m 41.28^s, -22\deg 32^\prime 33.7^{\prime\prime}$  &  Aug 19 at 16:47:19 (2.2) &   117  \\
\hline
\end{tabular}
\end{center}
\begin{flushleft}
$^a$ Observations were 2017, times in parentheses are days since the GW trigger.
\end{flushleft}
\end{table*}

\begin{table*}
\begin{center}
\begin{tabular}{ccc}
\hline
Pointing direction & Start time$^a$ & Exposure \\
(J2000) & (UTC) & (s) \\
\hline
$13^h 18^m 6.24^s, -27\deg 33^\prime 07.9^{\prime\prime}$  &  Aug 20 at 00:35:49 (2.5) &   117  \\
$13^h 18^m 16.32^s, -27\deg 51^\prime 55.1^{\prime\prime}$  &  Aug 20 at 00:38:32 (2.5) &   112  \\
$13^h 19^m 41.28^s, -27\deg 51^\prime 55.1^{\prime\prime}$  &  Aug 20 at 00:40:56 (2.5) &   117  \\
$13^h 18^m 27.84^s, -28\deg 10^\prime 42.2^{\prime\prime}$  &  Aug 20 at 00:43:22 (2.5) &   125  \\
$13^h 16^m 51.36^s, -27\deg 51^\prime 55.1^{\prime\prime}$  &  Aug 20 at 00:45:53 (2.5) &   125  \\
$12^h 43^m 27.84^s, -15\deg 39^\prime 16.6^{\prime\prime}$  &  Aug 20 at 01:54:17 (2.6) &   72  \\
\hline
\end{tabular}
\end{center}
\begin{flushleft}
$^a$ Observations were 2017, times in parentheses are days since the GW trigger.
\end{flushleft}
\end{table*}

\begin{table*}
\begin{center}
\begin{tabular}{ccc}
\hline
Pointing direction & Start time$^a$ & Exposure \\
(J2000) & (UTC) & (s) \\
\hline
$13^h 13^m 43.92^s, -27\deg 14^\prime 20.8^{\prime\prime}$  &  Aug 20 at 07:11:27 (2.8) &   117  \\
$13^h 13^m 37.44^s, -26\deg 55^\prime 33.6^{\prime\prime}$  &  Aug 20 at 07:13:52 (2.8) &   112  \\
$13^h 22^m 20.40^s, -27\deg 33^\prime 07.9^{\prime\prime}$  &  Aug 20 at 07:16:17 (2.8) &   122  \\
$13^h 36^m 15.84^s, -33\deg 30^\prime 03.2^{\prime\prime}$  &  Aug 20 at 07:18:50 (2.8) &   117  \\
$14^h 04^m 23.28^s, -48\deg 31^\prime 45.8^{\prime\prime}$  &  Aug 20 at 07:21:29 (2.8) &   117  \\
$14^h 19^m 33.84^s, -57\deg 55^\prime 19.9^{\prime\prime}$  &  Aug 20 at 07:24:19 (2.8) &   117  \\
\hline
\end{tabular}
\end{center}
\begin{flushleft}
$^a$ Observations were 2017, times in parentheses are days since the GW trigger.
\end{flushleft}
\end{table*}

\begin{table*}
\begin{center}
\begin{tabular}{ccc}
\hline
Pointing direction & Start time$^a$ & Exposure \\
(J2000) & (UTC) & (s) \\
\hline
$12^h 47^m 48.96^s, -18\deg 28^\prime 21.0^{\prime\prime}$  &  Aug 20 at 19:23:57 (3.3) &   22  \\
$12^h 49^m 8.16^s, -18\deg 28^\prime 21.0^{\prime\prime}$  &  Aug 20 at 19:26:42 (3.3) &   107  \\
$12^h 50^m 27.36^s, -18\deg 28^\prime 21.0^{\prime\prime}$  &  Aug 20 at 19:29:07 (3.3) &   112  \\
$12^h 51^m 46.56^s, -18\deg 28^\prime 21.0^{\prime\prime}$  &  Aug 20 at 19:31:32 (3.3) &   120  \\
$12^h 51^m 12.96^s, -18\deg 47^\prime 08.2^{\prime\prime}$  &  Aug 20 at 19:33:57 (3.3) &   115  \\
$12^h 52^m 0.00^s, -19\deg 05^\prime 55.3^{\prime\prime}$  &  Aug 20 at 19:36:22 (3.3) &   120  \\
\hline
\end{tabular}
\end{center}
\begin{flushleft}
$^a$ Observations were 2017, times in parentheses are days since the GW trigger.
\end{flushleft}
\end{table*}

\begin{table*}
\begin{center}
\begin{tabular}{ccc}
\hline
Pointing direction & Start time$^a$ & Exposure \\
(J2000) & (UTC) & (s) \\
\hline
$13^h 08^m 59.28^s, -21\deg 17^\prime 25.1^{\prime\prime}$  &  Aug 21 at 00:31:18 (3.5) &   120  \\
$13^h 08^m 39.84^s, -21\deg 36^\prime 12.2^{\prime\prime}$  &  Aug 21 at 00:33:47 (3.5) &   120  \\
$13^h 10^m 19.92^s, -21\deg 17^\prime 25.1^{\prime\prime}$  &  Aug 21 at 00:36:13 (3.5) &   120  \\
$13^h 10^m 40.56^s, -20\deg 58^\prime 37.9^{\prime\prime}$  &  Aug 21 at 00:38:44 (3.5) &   115  \\
$13^h 11^m 2.64^s, -20\deg 39^\prime 50.8^{\prime\prime}$  &  Aug 21 at 00:41:09 (3.5) &   107  \\
$13^h 38^m 32.64^s, -36\deg 37^\prime 54.8^{\prime\prime}$  &  Aug 21 at 04:03:31 (3.6) &   82  \\
\hline
\end{tabular}
\end{center}
\begin{flushleft}
$^a$ Observations were 2017, times in parentheses are days since the GW trigger.
\end{flushleft}
\end{table*}

\begin{table*}
\begin{center}
\begin{tabular}{ccc}
\hline
Pointing direction & Start time$^a$ & Exposure \\
(J2000) & (UTC) & (s) \\
\hline
$13^h 32^m 27.36^s, -33\deg 48^\prime 50.4^{\prime\prime}$  &  Aug 21 at 10:10:55 (3.9) &   120  \\
$13^h 30^m 56.88^s, -33\deg 48^\prime 50.4^{\prime\prime}$  &  Aug 21 at 10:13:20 (3.9) &   120  \\
$13^h 30^m 15.36^s, -33\deg 30^\prime 03.2^{\prime\prime}$  &  Aug 21 at 10:15:45 (3.9) &   115  \\
$13^h 29^m 35.52^s, -33\deg 11^\prime 16.4^{\prime\prime}$  &  Aug 21 at 10:18:10 (3.9) &   120  \\
$13^h 28^m 57.60^s, -32\deg 52^\prime 29.3^{\prime\prime}$  &  Aug 21 at 10:20:35 (3.9) &   120  \\
$13^h 28^m 5.76^s, -33\deg 11^\prime 16.4^{\prime\prime}$  &  Aug 21 at 10:23:00 (3.9) &   117  \\
\hline
\end{tabular}
\end{center}
\begin{flushleft}
$^a$ Observations were 2017, times in parentheses are days since the GW trigger.
\end{flushleft}
\end{table*}

\begin{table*}
\begin{center}
\begin{tabular}{ccc}
\hline
Pointing direction & Start time$^a$ & Exposure \\
(J2000) & (UTC) & (s) \\
\hline
$13^h 37^m 46.08^s, -33\deg 30^\prime 03.2^{\prime\prime}$  &  Aug 21 at 13:24:15 (4.0) &   117  \\
$13^h 38^m 29.28^s, -33\deg 48^\prime 50.4^{\prime\prime}$  &  Aug 21 at 13:26:40 (4.0) &   112  \\
$13^h 39^m 59.52^s, -33\deg 48^\prime 50.4^{\prime\prime}$  &  Aug 21 at 13:29:05 (4.0) &   117  \\
$13^h 40^m 46.32^s, -33\deg 30^\prime 03.2^{\prime\prime}$  &  Aug 21 at 13:31:30 (4.0) &   117  \\
$13^h 39^m 16.08^s, -33\deg 30^\prime 03.2^{\prime\prime}$  &  Aug 21 at 13:33:55 (4.0) &   115  \\
$13^h 40^m 6.24^s, -36\deg 37^\prime 54.8^{\prime\prime}$  &  Aug 21 at 13:36:20 (4.0) &   115  \\
\hline
\end{tabular}
\end{center}
\begin{flushleft}
$^a$ Observations were 2017, times in parentheses are days since the GW trigger.
\end{flushleft}
\end{table*}

\begin{table*}
\begin{center}
\begin{tabular}{ccc}
\hline
Pointing direction & Start time$^a$ & Exposure \\
(J2000) & (UTC) & (s) \\
\hline
$13^h 16^m 28.32^s, -24\deg 06^\prime 29.5^{\prime\prime}$  &  Aug 21 at 16:38:55 (4.2) &   125  \\
$13^h 16^m 22.56^s, -24\deg 25^\prime 16.7^{\prime\prime}$  &  Aug 21 at 16:41:25 (4.2) &   115  \\
$13^h 14^m 55.20^s, -24\deg 44^\prime 03.5^{\prime\prime}$  &  Aug 21 at 16:43:50 (4.2) &   122  \\
$13^h 29^m 15.36^s, -32\deg 14^\prime 55.0^{\prime\prime}$  &  Aug 21 at 16:46:21 (4.2) &   117  \\
$13^h 47^m 29.76^s, -39\deg 08^\prime 11.8^{\prime\prime}$  &  Aug 21 at 16:49:02 (4.2) &   117  \\
$14^h 08^m 23.04^s, -55\deg 25^\prime 03.0^{\prime\prime}$  &  Aug 21 at 16:51:42 (4.2) &   140  \\
\hline
\end{tabular}
\end{center}
\begin{flushleft}
$^a$ Observations were 2017, times in parentheses are days since the GW trigger.
\end{flushleft}
\end{table*}

\bibliography{ms_science}

\bibliographystyle{Science}

\end{document}